\documentclass{article}
\usepackage[numbers,super]{natbib}

\usepackage{PRIMEarxiv}
\usepackage{float}
\usepackage{amsmath}
\usepackage[utf8]{inputenc} 
\usepackage[T1]{fontenc}    
\usepackage{hyperref}       
\usepackage{url}            
\usepackage{booktabs}       
\usepackage{amsfonts}       
\usepackage{nicefrac}       
\usepackage{microtype}      
\usepackage{lipsum}
\usepackage{fancyhdr}       
\usepackage{graphicx}       
\usepackage{lineno}
\graphicspath{{media/}}     
\title{Advantages of Broadband Metalenses for Generalizable Image Classification}
\author{
  Yubo Zhang\textsuperscript{a} \\
  \AND
  Johannes Fr\"och\textsuperscript{a} \\
  \And 
  Jinlin Xiang\textsuperscript{a} \\
  \And
  Shane Colburn\textsuperscript{a} \\
  \And
  Myunghoo Lee\textsuperscript{a} \\
  \And
  Zhihao Zhou\textsuperscript{a} \\
  \And
  Minho Choi\textsuperscript{b} \\
  \And
  Eli Shlizerman\textsuperscript{a} \\
  \And 
  Arka Majumdar\textsuperscript{a*} \\
}

\begin{document}
\maketitle
\begin{center}
\textsuperscript{a}Department of Electrical \& Computer Engineering, University of Washington, Seattle, WA 98195, United States\\
\textsuperscript{b}Department of Electrical Engineering, Ulsan National Institute of Science and Technology (UNIST), Ulsan 44919, Republic of Korea \\
\textsuperscript{*}Email: arka@uw.edu
\end{center}

\begin{abstract}
Optical neural networks (ONNs) are gaining increasing attention to accelerate machine learning tasks. In particular, static meta-optical encoders designed for task-specific pre-processing have demonstrated orders of magnitude smaller energy consumption over purely digital counterparts, albeit at the cost of a slight degradation in classification accuracy. However, a lack of generalizability poses serious challenges for wide deployment of static meta-optical front-ends. Here, we investigate the utility of a single-layer metalens as a meta-optical encoder in ONNs for generalizable image classification. Specifically, we show that a visible-spectrum broadband metalens can achieve image classification accuracy comparable to high-end, sensor-limited optics and consistently outperforms the corresponding hyperboloid baseline across a wide range of sensor pixel sizes and digital backends. We further design an end-to-end optimized single-aperture metasurface for ImageNet classification and observe that the optimization tends to balance the modulation transfer function (MTF) across wavelengths within the sensor-detectable passband. Together, these observations suggest that the preservation of spatial-frequency information is an important factor influencing the performance of ONNs. Our results provide physical insight into the process of task-driven optical optimization and offer practical guidance for the design of high-performance ONNs and meta-optical encoders for generalizable computer vision tasks.

\end{abstract}
\keywords{Meta-Optical Encoder \and Optical Neural Networks (ONNs) \and End-to-end Optimization \and Modulation Transfer Function (MTF) \and Broadband Metalens \and Generalizable Image Classification}

\section{Introduction}

Diffractive optics leverage the high spatial bandwidth of light and wavefront engineering to precisely modulate spatial light intensity, and have been widely applied to tasks such as imaging\cite{JF_imaging,zhao2025endoscopy,zeng2025edof}, beam shaping\cite{Shane_beamshaping, capasso2014mo, nature2025beam_steering}, filtering\cite{goodam1992filtermatch} and holography\cite{Ozkan_holography, optica2024holography_encrypt}. With these physical capabilities, diffractive optical neural networks (ONNs)\cite{Zheng2022-zw, D2NN_Ozk, ozkan_DNN_nature, Iran2025programmable_DNN} have been shown to offer advantages in massive parallelism, rapid inference\cite{ozkan_optic_generative_nature}, and energy efficiency\cite{huang_photonic_adv, tsinghua2024onn_review}.
To date, research has shown that realizing fully optical neural network remains highly challenging---not simply because of limited nonlinear operations\cite{all_optical_nonlinearity2019optica} or programmability\cite{MONN_review2025programmable}, but also because achieving advantages over modern electronic processors requires carefully leveraging multiple optical features simultaneously while avoiding a number of fundamental pitfalls\cite{McMahon_review_2023}. A practical implementation of ONNs therefore adopts a hybrid optical-electronic architecture, where a static optical front-end preprocesses the input before feeding it into a digital back-end\cite{minho2025review}. Such systems offload part of the computational burden from the digital domain to optics, enabling faster inference and improved energy efficiency. 

Among various implementations of optical front-ends, meta-optics provides a compact and efficient solution\cite{Minho_cifar10}. Meta-optical encoders, built upon sub-wavelength-scale nano-structures, allow dense and parallel modulation of the amplitude, phase, and polarization of light, often achieving hundreds of optical operations on a single chip\cite{arka_mo_review, ms_array2024singleshot_ellipsometry, capasso2024singleshot}. Under incoherent illumination, the image on sensor is the convolution of the  object intensity and the incoherent point spread function (PSF) of the optical system\cite{princeton_spatial_varying, goodman1978incoherent, incoherent2023universal_linear}. Thus, by engineering the PSF of the meta-optics, we can implement different convolutions and effectively replace early convolutional layers in a neural network. Recent studies demonstrated that PSF-engineered meta-optical encoders, combined with advanced model compression techniques such as knowledge distillation\cite{Jinlin_KDtheory, zju2023knowledge_distillation}, can enable robust and energy-efficient ONNs for tasks such as classification\cite{Minho_cifar10} and segmentation\cite{jinlin_seg, another_segmentation_paper}. 


Although theoretical results for meta-optical encoder such as the universal approximation theorem and neural tangent kernel dynamics suggest that sufficiently wide neural layers can approximate deeper networks~\cite{jinlin_KDKincre}, most optical ONNs still exhibit reduced task performance compared to their purely digital counterparts in practice. This performance gap arises from both physical constraints and information loss in optical encoding. Optical implementations cannot realize arbitrary convolution kernels, and approximations such as the on-axis assumption can introduce deviations between design and experiment~\cite{NUS_spatial_varying}. At the same time, information lost during optical encoding irreversibly degrades the quality of the input features, limiting the achievable performance even when identical digital back-end architectures are used~\cite{huang_photonic_adv, Minho_cifar10, jinlin_seg}.

\begin{figure}[htbp]
    \centering
    \includegraphics[width=0.9\textwidth]{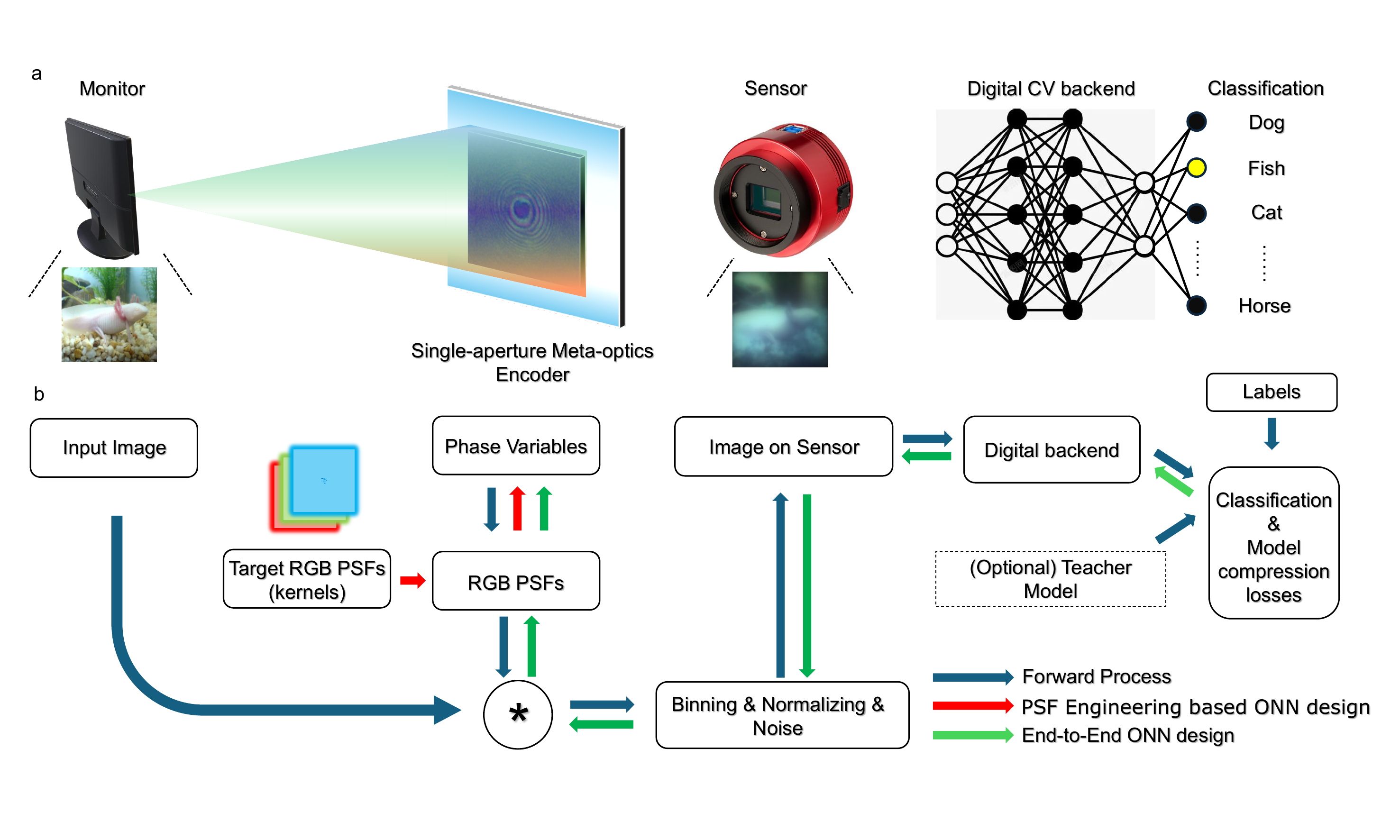}
    \caption{a. Schematic of a hybrid optical neural network (ONN) consisting of a single-aperture meta-optical encoder followed by a digital classification backend. An example image from ImageNet30 is encoded by the meta-optic, captured by the sensor, and processed by the downstream neural network for classification. b. Workflow of PSF engineering and end-to-end optimization in the hybrid ONN system. The differentiable forward propagation through the optical system and digital network is indicated by blue arrows. In the PSF-engineering approach (red arrows), the design proceeds in two steps: first, specifying a target PSF, which may be derived from convolutional neural network's filters or from optical design priors (like broadband optimization); second, inversely designing the meta-optic phase profile to realize the desired PSF. In contrast, the end-to-end design approach (green arrows) jointly optimizes the optical encoder and digital backend within a differentiable simulation pipeline, where sensor-plane signals are propagated through the network and the overall task loss is minimized via gradient descent.}
    \label{fig:1}
\end{figure}

One promising strategy to mitigate those limitation is end-to-end optimization\cite{shane_neuralnano, johannes_beating, Zinlin2024E2Edesign, e2e_framework2021design, stanford2018classification, tsinghua2025modesign} (see Fig \ref{fig:1}), in which the meta-optical front-end is modeled within a differentiable physical propagation framework and jointly trained with the digital back-end using gradient-based optimization. This approach enables the system to partially compensate for optical constraints and imperfections, often improving robustness and task performance. However, such designs are typically optimized for a specific dataset or digital backend, and therefore lack generalizability across different imaging conditions, sensor configurations, or downstream networks. This limited generalizability can hinder their broader deployment in practical computer vision systems.

To develop more generalizable optical encoders, it is desirable to analyze optical information preservation using physically interpretable metrics. While there are some information-theoretic tools to quantify representation efficiency in optical encoder systems\cite{lensless_imaging, laura_another2024information}, these approaches often lack direct physical interpretation in context of optical imaging. The modulation transfer function (MTF; Supporting Information: MTF definition), in contrast, provides an intuitive and physically grounded measure of how spatial-frequency information is transmitted through an optical system.

In this work, we investigate how the MTF of an optical encoder influences task performance in hybrid ONNs. 
Rather than pursuing state-of-the-art accuracy on a specific downstream task, our work aims to identify physically interpretable trends underlying the observed performance differences, with the goal of informing optical encoder designs that remain robust across different sensor configurations and digital backends.
Through experiments with fabricated centimeter-scale metalenses, we show that a broadband imaging–optimized metalens in visible spectrum consistently achieves higher classification accuracy than a hyperboloid metalens across multiple sensor pixel sizes and digital backend architectures. Interestingly, when the imaging system becomes limited by sensor resolution, the broadband metalens attains classification performance comparable to that of a high-end diffraction-limited refractive lens.

To understand the origin of this behavior, we further analyze end-to-end optimized meta-optical encoders in task-driven classification simulations. We observe that task-driven optimization systematically redistributes spatial-frequency transmission within the sensor-detectable passband, whose upper limit is set by the sensor Nyquist frequency determined by the pixel size, leading to an increased wavelength-averaged in-band MTF integral and a more balanced in-band MTF profile across wavelengths. This trend is consistent with the behavior observed in broadband imaging–optimized metalenses.

Together, these observations suggest a consistent frequency-domain behavior underlying generalizable image classification in hybrid ONNs: preserving spatial-frequency information within the sensor-detectable passband is closely associated with improved downstream task performance. This perspective offers a physically interpretable framework for analyzing and guiding the design of meta-optical encoders for general-purpose computer vision systems.

\section{Design Principle and Results}

\subsection{Hybrid ONN Setup and Optimization Framework}

We consider a hybrid ONN consisting of a single-aperture, rotationally symmetric meta-optic acting as the optical encoder, followed by a digital backend for ImageNet30 classification (a 30-class subset of ImageNet), as illustrated in Fig.~\ref{fig:1}a. The rotational symmetry is adopted to maintain generalizability and avoid task-specific spatial bias in the optical encoder.

As shown in Fig.~\ref{fig:1}b, typically the optical encoder can be designed using two different approaches: either (i) end-to-end optimization that directly maximizes the final classification accuracy, or (ii) PSF engineering, where the phase profile of the meta-optic is designed to produce a desired point spread function (PSF, whose Fourier transform defines the optical transfer function (OTF; see Supporting Information: MTF definition), with its normalized magnitude giving the MTF).

The end-to-end approach is conceptually straightforward, where a differentiable physical model of the optical system is integrated with the digital neural network and optimized through gradient descent. However, in practice this procedure requires joint optimization over both the optical encoder and the digital backend. Backpropagation through the entire hybrid system may lead to unstable training dynamics, or converge to suboptimal solutions. Furthermore, optical encoders obtained through task-specific end-to-end optimization may not generalize well across different sensor pixel sizes or digital backend architectures, as further discussed in Supporting Information: transferability analysis.

In contrast, broadband imaging metalenses are commonly designed using PSF engineering. In this approach, the phase profile of the meta-optic is optimized to maximize the wavelength-averaged MTF integral\cite{luocheng_broadband}. This optimization improves and balances spatial-frequency transmission across wavelengths. Notably, this design approach is agnostic to the downstream digital backend and sensor configuration, and typically exhibits faster and more stable optimization convergence.

Based on these two approaches, we perform both experimental and simulation studies.

(1) Experimental setup.
Experimentally, we use a static 1-cm-aperture broadband metalens operating in the visible spectrum as the optical encoder, based on the design reported in \cite{johannes_beating}. In the measurement system, a computer monitor displays images from ImageNet30 with incoherent light. The light is further captured by a camera (ZWO ASI 678MC) after passing through the metalens. The recorded sensor measurements are then used as inputs to the digital neural network backend for inference. For comparison, we repeat the same procedure using a corresponding 1-cm-aperture hyperboloid metalens as the baseline.

(2) Simulation setup.
In simulations, we perform end-to-end optimization of a 0.5-mm-aperture meta-optic encoder jointly with a digital backend for image classification. All components of the system, including optical propagation and sensor sampling, are modeled within a differentiable framework to enable gradient-based optimization. The smaller aperture is chosen for computational efficiency, since larger apertures require significantly more scattering elements and substantially increase the optimization cost.

Although the experimental and simulated meta-optics differ in aperture size, the two studies provide physical insights into how the frequency-domain behavior of optical encoders relates to downstream ONN task performance and how task-driven optical optimization influences the MTF within the sensor-detectable passband.

\subsection{Hyperboloid Metalens Baseline: Motivation and Limitations}

For both setups, a hyperboloid lens operating at the green wavelength $\lambda_0$ is used as the baseline. The hyperboloid serves as a natural reference for evaluating the end-to-end optimized design, representing a conventional, analytically defined phase profile that focuses light according to geometrical optics principles.

\[
\phi(x,y) = \frac{2\pi}{\lambda_0} \left( f - \sqrt{x^{2} + y^{2} + f^{2}}  \right)
\]

where $f$ denotes the focal length. This phase profile corresponds to the planar implementation of a conventional refractive lens and is widely used as a standard design in meta-optics. 

Such a lens is optimized only for diffraction-limited focusing at the design wavelength $\lambda_0$. Away from this wavelength, chromatic dispersion and phase wrapping introduce distortions that degrade imaging performance. The hyperboloid lens does not incorporate task-specific optimization and serves as a suitable baseline for evaluating the benefits of task-driven end-to-end optimization or broadband MTF optimization.

To ensure a fair comparison, both the broadband metalens used in the experiment and the end-to-end optimized metalens used in simulation are constrained to be rotationally symmetric (Supporting Information: end-to-end optimization methods), implemented using a polynomial parameterization of the phase profile. This constraint reduces the number of optimization parameters, improves numerical stability, and promotes better task generalizability.

\begin{figure}[htbp]
    \centering
    \includegraphics[width=0.9\textwidth]{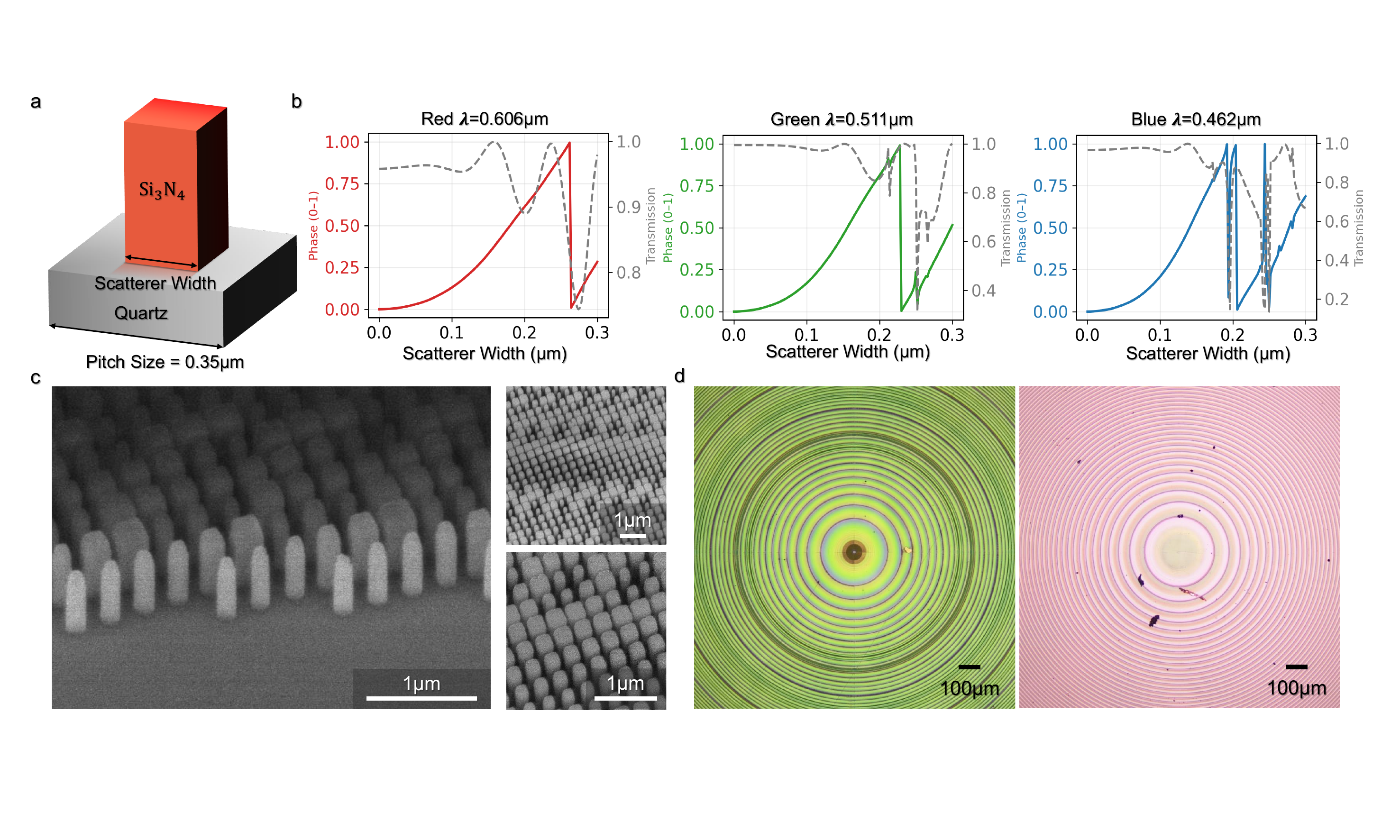}
    \caption{a. Schematic of a single meta-scatterer in the $\mathrm{Si_3N_4}$-on-quartz meta-optics.
    b. RCWA(rigorous coupled-wave analysis) simulated normalized phase and amplitude responses of single meta-scatterer at red (606 nm), green (511 nm), and blue (462 nm) wavelengths.
    c. Scanning electron microscope (SEM) images of the fabricated broadband meta-optics.
    d. Optical microscope images of the fabricated broadband meta-optics (left) and the hyperboloid baseline (right).}
    \label{fig:2}
\end{figure}

\subsection{Characterization of 1-cm-aperture Broadband Metalens}

The 1-cm-aperture meta-lenses (broadband and hyperboloid) were fabricated on a $\mathrm{Si_3N_4}$-on-quartz wafer, as illustrated in Fig.~\ref{fig:2}a. It was designed for broadband imaging and we experimentally verified that the broadband metalens results in a balanced spatial-frequency response across the visible spectrum. Details of the fabrication process and experimental materials are provided in the Supplement information (Supporting Information: materials and fabrication details).
The optical responses of the $\mathrm{Si_3N_4}$ meta-atoms were simulated using rigorous coupled-wave analysis (RCWA), as shown in Fig.~\ref{fig:2}b for scatterers designed at red (606 nm), green (511 nm), and blue (462 nm) wavelengths. Scanning electron micrographs and optical microscope images of the fabricated broadband and hyperboloid metalenses are shown in Figs.~\ref{fig:2}c and~\ref{fig:2}d.

\begin{figure}[htbp]
    \centering
    \includegraphics[width=0.9\textwidth]{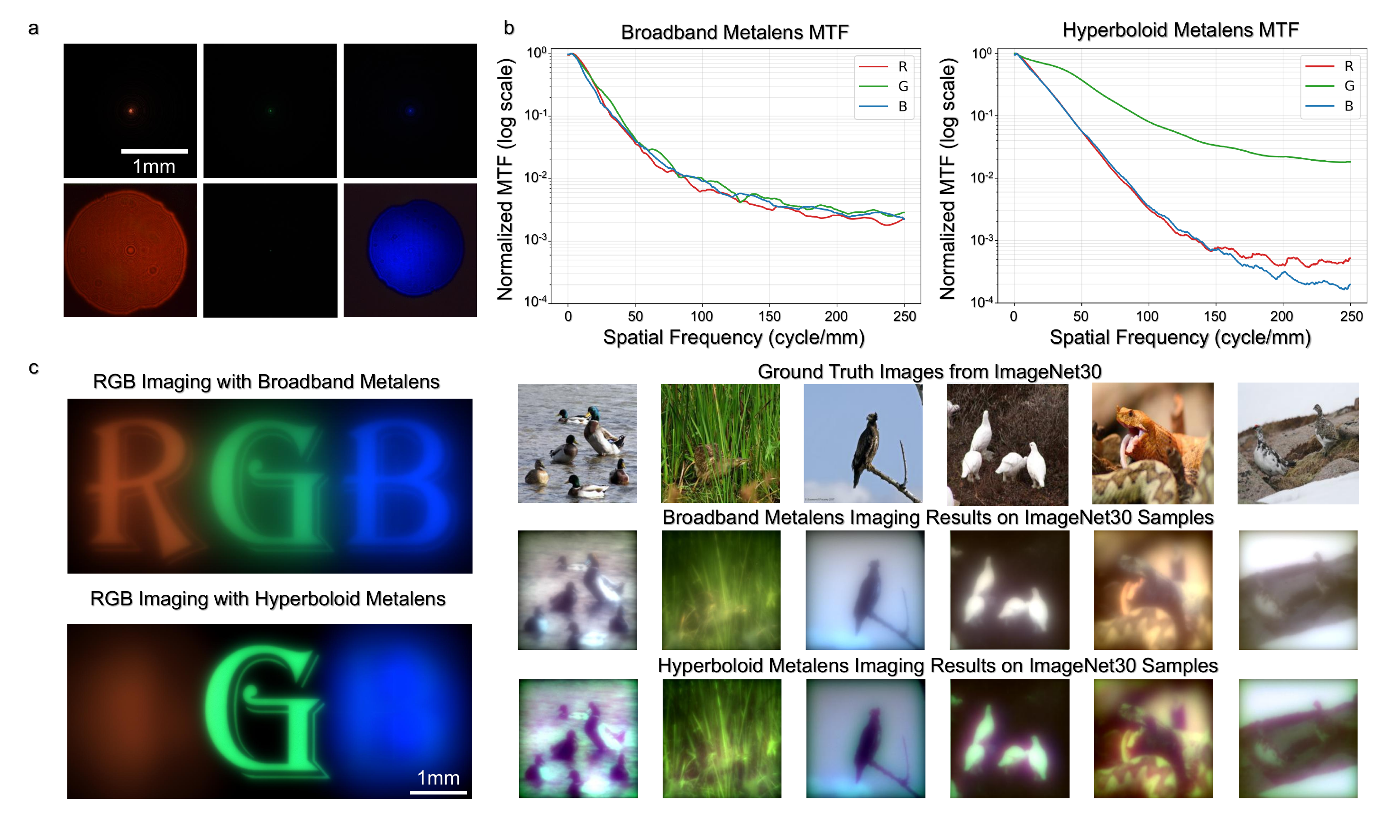}
    \caption{a. RGB point spread functions (PSFs) of the broadband metalens (top) and the hyperboloid metalens (bottom). b. Log-scaled modulation transfer functions (MTFs) of the broadband metalens (left) and the hyperboloid metalens (right). c. (left) broadband(up)/hyperboloid(down) metalens imaging results of an RGB pattern. (right) Captured images of ImageNet30 by broadband/hyperboloid metalens on sensor.}
    \label{fig:3}
\end{figure}

Figure~\ref{fig:3}a shows the RGB PSFs of the two fabricated 1-cm-aperture meta-optics. The broadband metalens maintains compact PSFs across the red, green, and blue wavelengths, indicating balanced chromatic performance. In contrast, the hyperboloid metalens tightly focuses only the green wavelength, while the red and blue focal spots are noticeably defocused. This chromatic defocus results in spatially broadened PSFs at the red and blue wavelengths and degrades its color-imaging performance.

Figure~\ref{fig:3}b presents the experimentally measured MTF curves plotted on a log scale. The MTF is defined as the normalized magnitude of the optical transfer function (OTF), characterizing how different spatial frequencies are transmitted by the optical system. Under incoherent illumination, the MTF provides a convenient frequency-domain description of the imaging performance. The horizontal axis, expressed in cycles per millimeter, represents the spatial frequency up to the Nyquist limit imposed by the sensor sampling (2 µm pixel pitch).

For the hyperboloid metalens (right), only the green channel exhibits high MTF values across most spatial frequencies, while the red and blue channels experience strong attenuation due to chromatic defocus. In contrast, the broadband metalens (left) shows substantially improved MTF responses in the red and blue channels, resulting in a more spectrally balanced frequency response and a higher wavelength-averaged MTF within the sensor-detectable band.

The two metalenses exhibit distinct imaging characteristics, as shown in Fig.~\ref{fig:3}c. The left panel shows single-wavelength imaging of RGB letter patterns, while the right panel presents broadband color imaging results. The hyperboloid metalens produces sharper edges and higher local contrast at the design wavelength due to well-corrected on-axis focusing, but suffers from significant chromatic imbalance, with the red and blue channels noticeably blurred. In contrast, the broadband metalens provides more balanced image sharpness and color fidelity across the three channels, yielding slightly softer edges but a more uniform spectral response. As we show later, this balanced frequency-domain behavior is associated with improved performance in downstream computer-vision tasks.

\subsection{Advantage of the Fabricated Broadband Metalens for Generalizable Image Classification}

In this section, we evaluate ImageNet30 classification performance using the fabricated 1-cm broadband metalens as the optical encoder and compare it with a hyperboloid metalens across different digital backends and sensor pixel sizes. We also include the digital ground-truth as a contextual reference, which serves as a proxy for ideal conventional imaging performance from high-end optics.

\begin{figure}[htbp]
    \centering
    \includegraphics[width=1.0\textwidth]{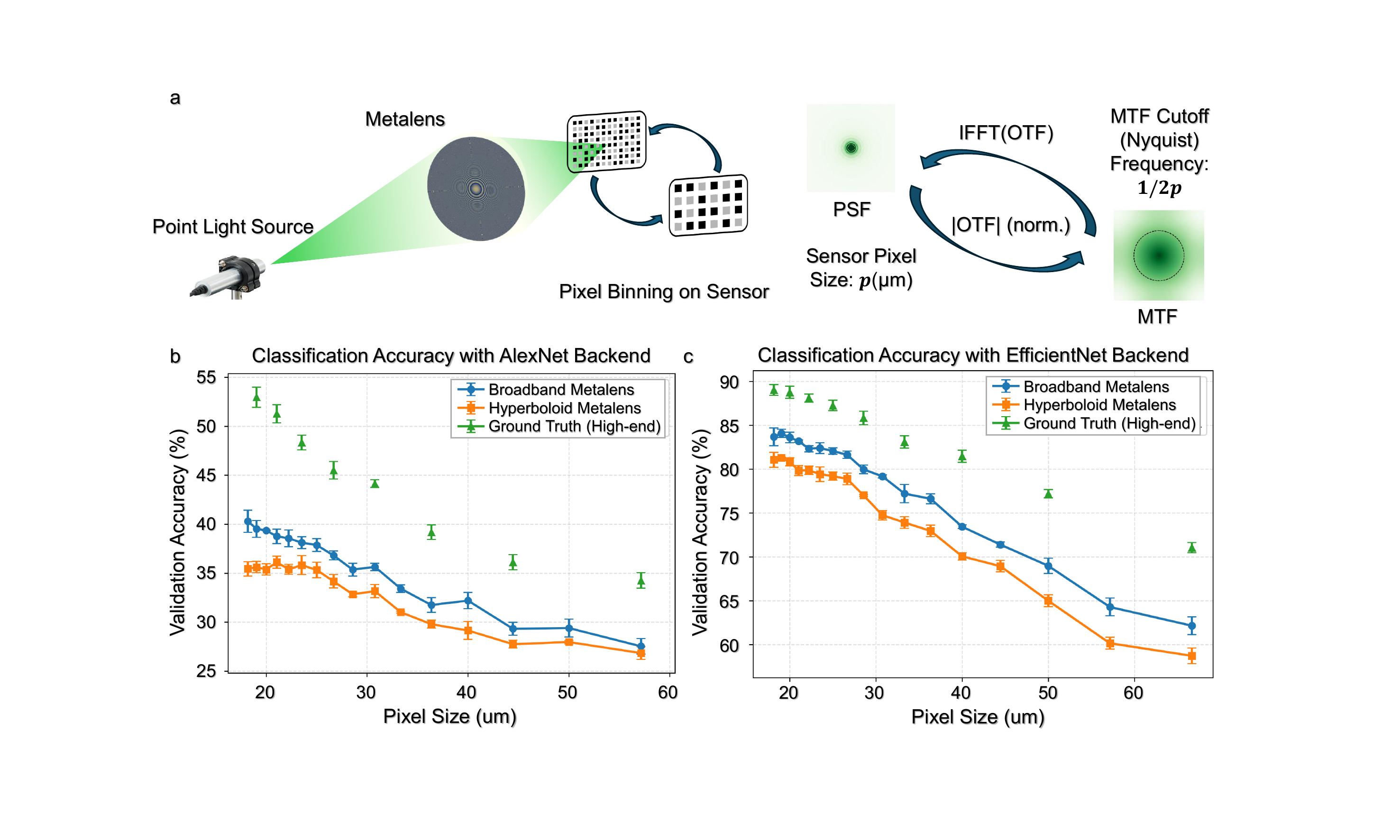}
    \caption{a. Effect of sensor pixel size on the detectable spatial-frequency passband of the imaging system. Increasing the pixel size spatially averages the incident intensity, smoothing fine PSF variations and suppressing high spatial frequencies. This operation is equivalent to a low-pass filtering in the frequency domain, reducing the system MTF cutoff frequency (1/2p). As a result, sensor sampling defines the spatial-frequency passband that can be transmitted to the digital backend. b\&c. Classification accuracy on ImageNet30 as a function of sensor pixel size, comparing AlexNet (b) and EfficientNet (c) digital backends. Results are shown for the digital ground-truth (proxy high-end optics), the broadband metalens, and the hyperboloid metalens. Error bars denote the standard deviation across five runs with different random seeds. Solid lines are included as visual guides and do not represent theoretical fits.
    }
    \label{fig:4}
\end{figure}

In hybrid ONNs, the sensor pixel size determines the effective sampling resolution and therefore limits the highest spatial frequency that can be captured. As illustrated in Fig.~\ref{fig:4}a, larger pixels average the incident optical intensity over a wider area, suppressing high spatial-frequency variations of the optical field within each pixel. This spatial averaging effectively reduces the system MTF cutoff frequency and narrows the spatial-frequency bandwidth available to the downstream digital network.

Figures~\ref{fig:4}b and~\ref{fig:4}c show classification results using AlexNet and  EfficientNet backends, representing digital models with different capacities. Each plot reports classification accuracy as a function of sensor pixel size for the digital ground-truth (proxy high-end optics), the broadband metalens, and the hyperboloid metalens. A consistent trend is observed: increasing the sensor pixel size reduces classification accuracy for both optical systems due to the loss of high-frequency information. Conversely, higher sensor resolution preserves finer spatial details and generally improves ONN performance.

Across all sensor pixel sizes, the broadband metalens consistently outperforms the hyperboloid design. The magnitude of this advantage depends on the digital backend. With the simpler AlexNet, the performance gap between the two optical encoders is smaller and both systems show larger degradation relative to the fully digital reference. In contrast, the higher-capacity EfficientNet better utilizes the spatial-frequency content transmitted by the broadband metalens, maintaining a clear advantage across resolutions.

Importantly, we observe that a broadband metalens combined with a higher-resolution sensor can achieve higher classification accuracy than the ground-truth system operating with coarse sampling. This comparison highlights the importance of considering the optical front end and sensor sampling jointly when evaluating hybrid ONN systems. 
More generally, in sensor-limited regimes where large pixel sizes impose a low-pass constraint on the captured spatial frequencies for low-power operation, the advantage of high-end optics over metalens is largely reduced as illustrated in Supporting Information: classification performance under large-pixel sensor regimes. In such scenarios, a compact broadband metalens can provide comparable task performance while offering potential benefits in system size and weight.

These observations suggest that the spatial-frequency content preserved by the optical encoder within the sensor-detectable band is an important factor in determining downstream classification performance.
The consistent advantage of the broadband metalens across pixel sizes and backend architectures is therefore better understood in terms of improved frequency-domain information preservation rather than spectral coverage alone. Although the optical layer contains far fewer trainable parameters than the digital backend, it governs how much spatial-frequency information is transmitted to the network. As a result, the design of the optical front end strongly influences the information available for the digital backend to exploit.

\begin{figure}[htbp]
    \centering
    \includegraphics[width=1.0\textwidth]{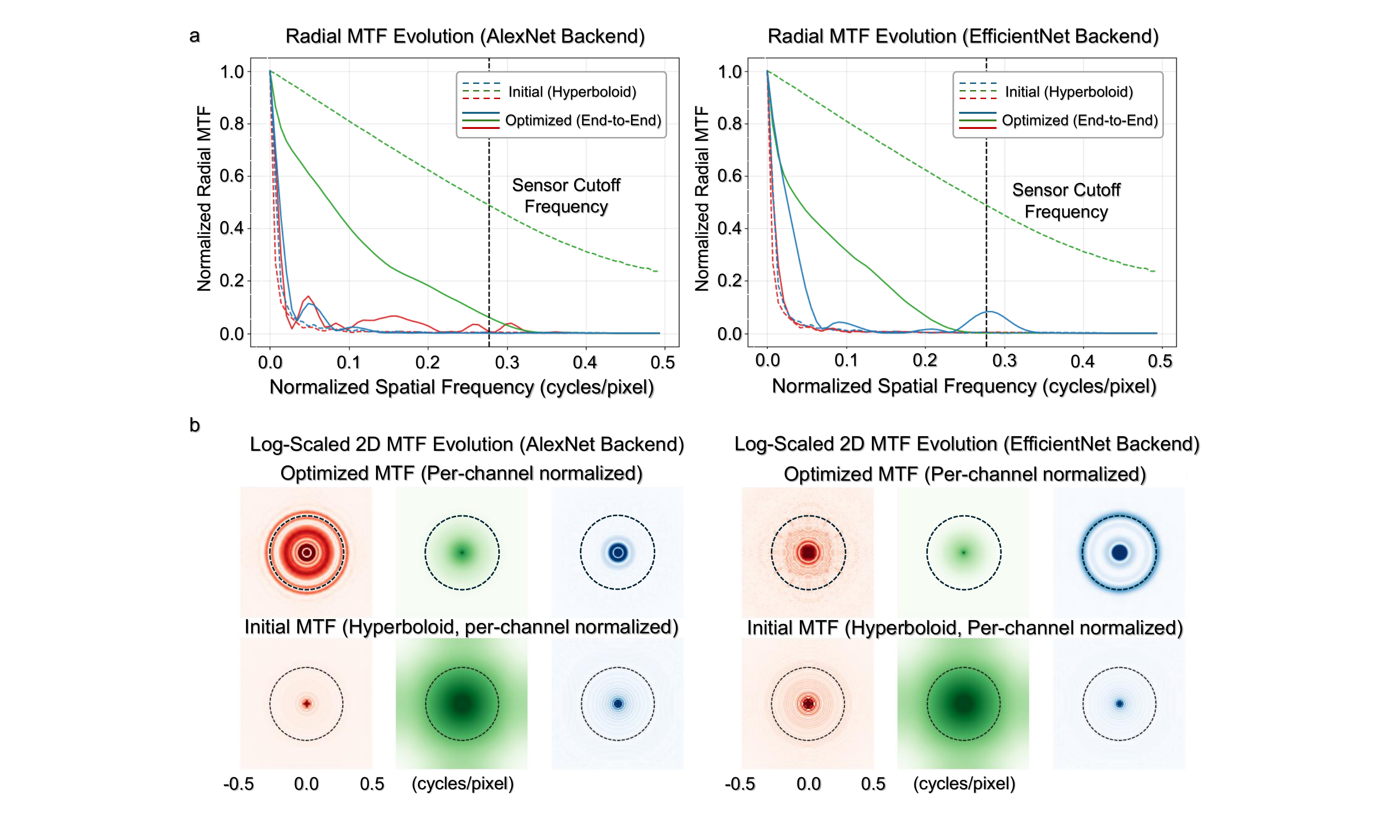}
    \caption{a. Radial MTF evolution of the end-to-end optimized metalens with AlexNet and EfficientNet backends. The dashed and solid curves correspond to the initial hyperboloid phase and the end-to-end optimized phase, respectively. The spatial-frequency axis is normalized to the Nyquist frequency, and the sensor cutoff frequency determined by the pixel size is indicated by the vertical dashed line. After optimization, the RGB MTF profiles become more balanced, and a larger fraction of spectral transmission is redistributed into the sensor-detectable band. b. Corresponding log-scaled 2D MTF distributions for each RGB channel, independently normalized. These heatmaps represent the full spatial-frequency distributions underlying the radial profiles in (a), highlighting the redistribution of in-band MTF across channels after optimization.}
    \label{fig:5}
\end{figure}

\subsection{MTF Evolution of Simulated End-to-End Optimized Metalenses}

To complement the experimental results, we simulate meta-optics optimized directly for classification accuracy via end-to-end training. Unlike the fabricated 1-cm broadband metalens, the simulated meta-optics are optimized with the explicit objective of maximizing classification performance.

For computational efficiency, we use a smaller-aperture simulated meta-optic and jointly optimize it with different digital neural-network backends (Supporting Information: MTF evolution across digital backends). The sensor pixel size was also varied to investigate how the optimization adapts to different sensor-imposed MTF cutoff frequencies.

The wavelength-dependent phase response was approximated using polynomial expansions (Supporting Information: end-to-end optimization methods), and light propagation was modeled using a differentiable angular spectrum method to enable gradient-based optimization. During training, separate learning rates were assigned to the optical phase variables and the digital-network parameters to ensure stable convergence.

The meta-optic was initialized from a hyperboloid phase profile. After convergence, the end-to-end optimized model achieved improved classification accuracy relative to the non-optimized baseline. Representative radial MTF curves and corresponding heatmaps are shown in Figs.~\ref{fig:5}a and~\ref{fig:5}b.

Compared with the initial hyperboloid phase, the end-to-end optimization exhibits two key trends. First, the integrated MTF areas of the RGB channels become more balanced, reflecting the need for consistent spectral information in dataset images. Second, the optimization redistributes the spectral MTF such that more transmission is concentrated within the sensor-detectable band. In particular, the initially green-dominated response becomes less spread into frequencies beyond the sensor-imposed cutoff, while the red and blue channels are enhanced primarily within the in-band region, with little corresponding increase outside the detectable range.

Notably, these trends closely mirror those observed in the experimentally fabricated broadband metalens. Although the two designs are optimized for different objectives—-broadband imaging and classification accuracy—-both exhibit improved spatial-frequency transmission within the sensor-detectable band. This consistency suggests that the frequency-domain behavior is likely an important factor in more generalizable classification performance in hybrid ONNs.


\begin{figure}[htbp]
    \centering
    \includegraphics[width=1.0\textwidth]{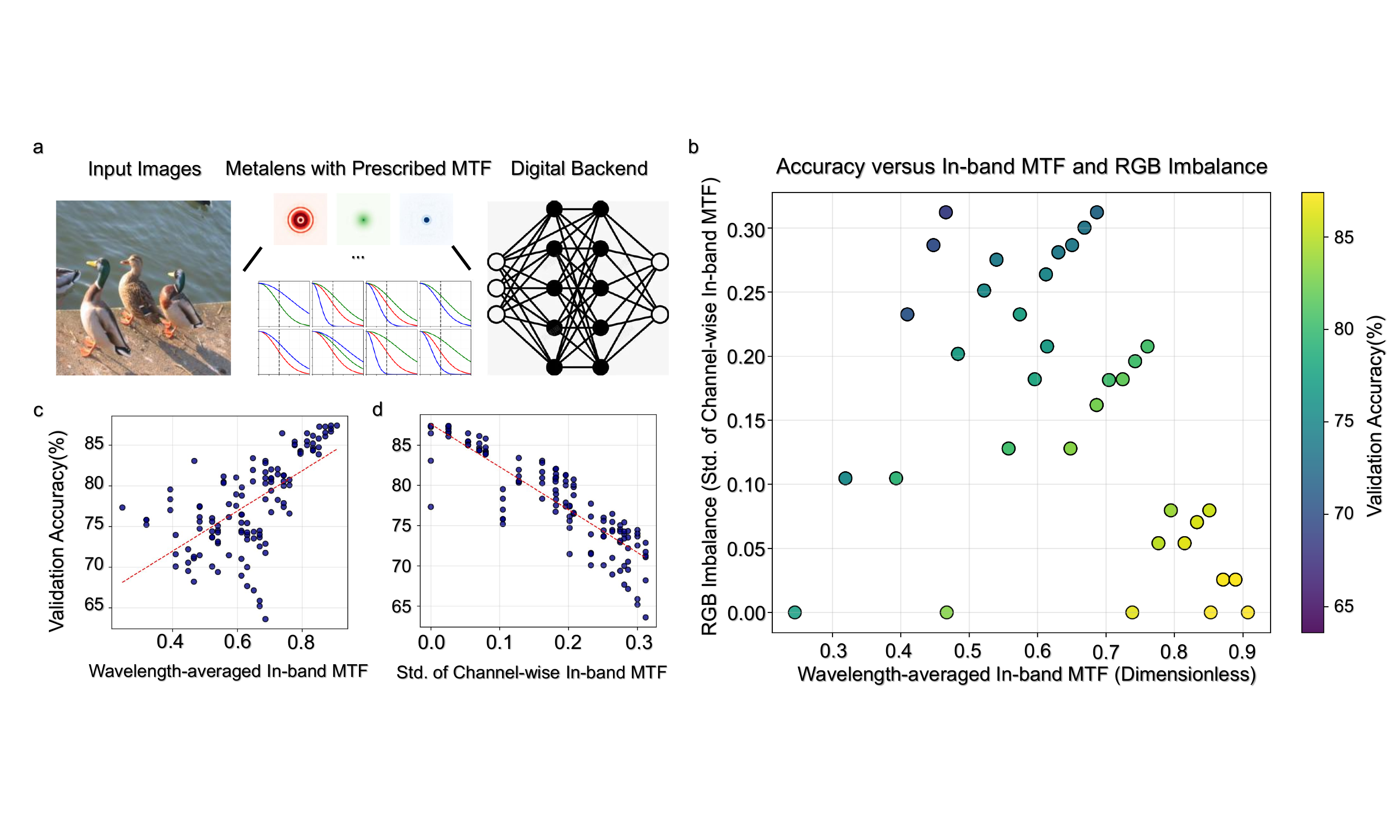}
    \caption{a. Illustration of the controlled synthetic-MTF framework. Input images are filtered by prescribed RGB optical transfer functions that emulate meta-optical encoders with engineered frequency responses, and the resulting images are processed by a fixed digital backend.
    b. Validation accuracy as a function of wavelength-averaged in-band MTF and RGB imbalance. Each point corresponds to a synthesized RGB frequency-response configuration. Higher accuracy is associated with increased in-band MTF transmission and reduced cross-channel imbalance.
    c. Validation accuracy versus wavelength-averaged in-band MTF, showing a positive correlation(Pearson correlation coefficient r=0.632) between preserved in-band spatial-frequency transmission and task performance.
    d. Validation accuracy versus RGB imbalance (standard deviation of in-band MTF across channels), showing that strong spectral imbalance degrades performance(r=-0.867).
}
    \label{fig:6}
\end{figure}

\subsection{Controlled Validation with Synthesized MTF Profiles}

To further isolate the role of frequency-domain transmission in classification performance, we conduct a controlled study using synthesized MTF profiles, as illustrated in Fig.~\ref{fig:6}. Instead of optimizing a physical meta-optic, we directly impose prescribed optical transfer functions (OTFs) in the Fourier domain, enabling independent control over spatial-frequency transmission and cross-channel balance.

Specifically, RGB images from ImageNet30 with resolution $224\times224$ are filtered channel-wise by radially symmetric frequency responses and then processed by an identical digital backend (EfficientNet). To emulate sensor-limited imaging, we define an in-band cutoff frequency $f_0$ (in cycles/pixel) and evaluate all MTF-based quantities within the region $f_r \leq f_0$. By systematically varying the bandwidth of each color channel, we generate $125$ synthesized RGB (See Supporting Information: synthesized MTF experiments for details) frequency-response configurations that span a wide range of in-band transmission and cross-channel imbalance.

As shown in Fig.~\ref{fig:6}b, validation accuracy is jointly influenced by two key factors: the amount of spatial-frequency transmission within the sensor-detectable band and its balance across color channels. Higher accuracy is generally associated with increased wavelength-averaged in-band MTF and reduced RGB imbalance. This trend is further supported by the one-dimensional correlations in Fig.~\ref{fig:6}c–d, where accuracy increases with in-band MTF (Pearson correlation coefficient $r=0.632$) and decreases with cross-channel imbalance ($r=-0.867$).

These controlled experiments help isolate the effect of frequency-domain transmission from other optical and optimization factors, suggesting that downstream performance is closely related to in-band spatial-frequency transmission and its spectral distribution. We further examine additional MTF-related metrics in Supporting Information: synthesized MTF experiments and additional metrics and observe consistent trends with respect to classification accuracy.

\section{Conclusion}

In this work, we examined hybrid ONNs for image classification from a frequency-domain perspective. Across fabricated devices, end-to-end optical--digital co-optimization, and controlled synthetic-MTF validation, our results consistently indicate that downstream classification performance is closely related to how much spatial-frequency information is preserved within the sensor-detectable band and how evenly that information is distributed across color channels.

Experimentally, a broadband metalens exhibits a more balanced RGB MTF than a conventional hyperboloid metalens and correspondingly achieves higher classification accuracy across multiple digital backends and sensor pixel sizes. Complementary end-to-end simulations reveal similar optimization trends, with task-driven designs redistributing spatial-frequency transmission toward the sensor-accessible band while improving spectral balance. Controlled synthetic-MTF experiments further support this interpretation by showing that accuracy is most strongly associated with wavelength-averaged in-band MTF and RGB balance, while alternative MTF-related descriptors exhibit broadly consistent but weaker trends.

Taken together, these results suggest that in-band spatial-frequency transmission provides a useful and physically interpretable perspective for understanding task-driven optical optimization in hybrid ONNs. More broadly, this framework offers practical guidance for the design and deployment of meta-optical encoders for generalizable computer vision systems, particularly in sensor-limited regimes where compact meta-optics can remain competitive.

\section*{Supporting Information}

Materials and fabrication details; definition of MTF and its relation to PSF and sensor sampling; experimental setup and measurement procedures; additional imaging results; classification performance under large-pixel sensor regimes; feature-space clustering analysis; end-to-end optimization methods; transferability analysis; MTF evolution across digital backends; synthesized MTF experiments and additional metrics.

\section*{Funding Sources}

The research is supported by National Science Foundation (EFRI-BRAID-2223495). Part of this work was conducted at the Washington Nanofabrication Facility/ Molecular Analysis Facility, a National Nanotechnology Coordinated Infrastructure (NNCI) site at the University of Washington with partial support from the National Science Foundation via awards NNCI-1542101 and NNCI-2025489.

M. Choi was supported by Institute of Information \& communications Technology Planning \& Evaluation(IITP) grant funded by the Korea government(MSIT) (No.RS-2020-II201336, Artificial Intelligence graduate school support(UNIST)).

\section*{Acknowledgments}

The authors acknowledge the Washington Nanofabrication Facility for device fabrication support.

\bibliography{references}  

@article{luocheng_broadband,
  title={Broadband thermal imaging using meta-optics},
  author={Huang, Luocheng and Han, Zheyi and Wirth-Singh, Anna and Saragadam, Vishwanath and Mukherjee, Saswata and Fr{\"o}ch, Johannes E and Tanguy, Quentin AA and Rollag, Joshua and Gibson, Ricky and Hendrickson, Joshua R and others},
  journal={Nature Communications},
  volume={15},
  number={1},
  pages={1662},
  year={2024},
  publisher={Nature Publishing Group UK London}
}

@article{JF_imaging,
  title = {Full color visible imaging with crystalline silicon meta-optics},
  volume = {14},
  ISSN = {2047-7538},
  url = {http://dx.doi.org/10.1038/s41377-025-01888-w},
  DOI = {10.1038/s41377-025-01888-w},
  number = {1},
  journal = {Light: Science \& Applications},
  publisher = {Springer Science and Business Media LLC},
  author = {Fr\"{o}ch,  Johannes E. and Huang,  Luocheng and Zhou,  Zhihao and Tara,  Virat and Fang,  Zhuoran and Colburn,  Shane and Zhan,  Alan and Choi,  Minho and Manna,  Arnab and Tang,  Andrew and Han,  Zheyi and B\"{o}hringer,  Karl F. and Majumdar,  Arka},
  year = {2025},
  month = jun 
}

@article{Ozkan_holography,
  title = {Computer-Free,  All-Optical Reconstruction of Holograms Using Diffractive Networks},
  volume = {8},
  ISSN = {2330-4022},
  url = {http://dx.doi.org/10.1021/acsphotonics.1c01365},
  DOI = {10.1021/acsphotonics.1c01365},
  number = {11},
  journal = {ACS Photonics},
  publisher = {American Chemical Society (ACS)},
  author = {Sakib Rahman,  Md Sadman and Ozcan,  Aydogan},
  year = {2021},
  month = oct,
  pages = {3375–3384}
}

@article{Shane_beamshaping,
  title = {Metasurface Generation of Paired Accelerating and Rotating Optical Beams for Passive Ranging and Scene Reconstruction},
  volume = {7},
  ISSN = {2330-4022},
  url = {http://dx.doi.org/10.1021/acsphotonics.0c00354},
  DOI = {10.1021/acsphotonics.0c00354},
  number = {6},
  journal = {ACS Photonics},
  publisher = {American Chemical Society (ACS)},
  author = {Colburn,  Shane and Majumdar,  Arka},
  year = {2020},
  month = may,
  pages = {1529–1536}
}

@article{huang_photonic_adv,
  title={Photonic advantage of optical encoders},
  author={Huang, Luocheng and Tanguy, Quentin AA and Fr{\"o}ch, Johannes E and Mukherjee, Saswata and B{\"o}hringer, Karl F and Majumdar, Arka},
  journal={Nanophotonics},
  volume={13},
  number={7},
  pages={1191--1196},
  year={2024},
  publisher={De Gruyter}
}

@ARTICLE{Zheng2022-zw,
  title     = "Meta-optic accelerators for object classifiers",
  author    = "Zheng, Hanyu and Liu, Quan and Zhou, You and Kravchenko, Ivan I
               and Huo, Yuankai and Valentine, Jason",
  journal   = "Sci. Adv.",
  publisher = "American Association for the Advancement of Science (AAAS)",
  volume    =  8,
  number    =  30,
  pages     = "eabo6410",
  month     =  jul,
  year      =  2022,
}

@ARTICLE{D2NN_Ozk,
  author={Mengu, Deniz and Luo, Yi and Rivenson, Yair and Ozcan, Aydogan},
  journal={IEEE Journal of Selected Topics in Quantum Electronics}, 
  title={Analysis of Diffractive Optical Neural Networks and Their Integration With Electronic Neural Networks}, 
  year={2020},
  volume={26},
  number={1},
  pages={1-14},
  doi={10.1109/JSTQE.2019.2921376}}

@ARTICLE{Minho_cifar10,
  title     = "Transferable polychromatic optical encoder for neural networks",
  author    = "Choi, Minho and Xiang, Jinlin and Wirth-Singh, Anna and Baek,
               Seung-Hwan and Shlizerman, Eli and Majumdar, Arka",
  journal   = "Nat. Commun.",
  publisher = "Springer Science and Business Media LLC",
  volume    =  16,
  number    =  1,
  pages     = "5623",
  month     =  jul,
  year      =  2025,
  copyright = "https://creativecommons.org/licenses/by-nc-nd/4.0",
  language  = "en"
}

@article{jinlin_seg,
  title={Neural Tangent Knowledge Distillation for Optical Convolutional Networks},
  author={Xiang, Jinlin and Choi, Minho and Zhang, Yubo and Zhou, Zhihao and Majumdar, Arka and Shlizerman, Eli},
  journal={arXiv:2508.08421},
  year={2025}
}

@article{arka_mo_review,
  title={The second optical metasurface revolution: moving from science to technology},
  author={Brongersma, Mark L and Pala, Ragip A and Altug, Hatice and Capasso, Federico and Chen, Wei Ting and Majumdar, Arka and Atwater, Harry A},
  journal={Nature Reviews Electrical Engineering},
  volume={2},
  number={2},
  pages={125--143},
  year={2025},
  publisher={Nature Publishing Group UK London}
}

@article{princeton_spatial_varying,
  title={Spatially varying nanophotonic neural networks},
  author={Wei, Kaixuan and Li, Xiao and Froech, Johannes and Chakravarthula, Praneeth and Whitehead, James and Tseng, Ethan and Majumdar, Arka and Heide, Felix},
  journal={Science Advances},
  volume={10},
  number={45},
  pages={eadp0391},
  year={2024},
  publisher={American Association for the Advancement of Science}
}

@article{Jinlin_KDtheory,
  title={Knowledge distillation circumvents nonlinearity for optical convolutional neural networks},
  author={Xiang, Jinlin and Colburn, Shane and Majumdar, Arka and Shlizerman, Eli},
  journal={Applied Optics},
  volume={61},
  number={9},
  pages={2173--2183},
  year={2022},
  publisher={OSA}
}

@article{jinlin_KDKincre,
  title={Tkil: tangent kernel approach for class balanced incremental learning},
  author={Xiang, Jinlin and Shlizerman, Eli},
  journal={arXiv:2206.08492},
  year={2022}
}

@article{lensless_imaging,
  title={Designing lensless imaging systems to maximize information capture},
  author={Kabuli, Leyla A and Pinkard, Henry and Markley, Eric and Hung, Clara S and Waller, Laura},
  journal={arXiv:2506.08513},
  year={2025}
}

@article{minho2025review,
  title={Free-space optical encoder for computer vision},
  author={Choi, Minho and Majumdar, Arka},
  journal={npj Nanophotonics},
  volume={2},
  number={1},
  pages={36},
  year={2025},
  publisher={Nature Publishing Group UK London}
}

@article{NUS_spatial_varying,
  title={EigenCWD: a spatially varying deconvolution algorithm for single metalens imaging},
  author={Yeo, Joel and Duane Loh, N and Paniagua-Dominguez, Ramon and Kuznetsov, Arseniy I},
  journal={Optics Express},
  volume={33},
  number={13},
  pages={28481--28492},
  year={2025},
  publisher={Optica Publishing Group}
}

@article{shane_neuralnano,
  title={Neural nano-optics for high-quality thin lens imaging},
  author={Tseng, Ethan and Colburn, Shane and Whitehead, James and Huang, Luocheng and Baek, Seung-Hwan and Majumdar, Arka and Heide, Felix},
  journal={Nature Communications},
  volume={12},
  number={1},
  pages={6493},
  year={2021},
  publisher={Nature Publishing Group UK London}
}

@article{johannes_beating,
  title={Beating spectral bandwidth limits for large aperture broadband nano-optics},
  author={Fr{\"o}ch, Johannes E and Chakravarthula, Praneeth and Sun, Jipeng and Tseng, Ethan and Colburn, Shane and Zhan, Alan and Miller, Forrest and Wirth-Singh, Anna and Tanguy, Quentin AA and Han, Zheyi and others},
  journal={Nature Communications},
  volume={16},
  number={1},
  pages={3025},
  year={2025},
  publisher={Nature Publishing Group UK London}
}

@article{McMahon_review_2023,
   title={The physics of optical computing},
   volume={5},
   ISSN={2522-5820},
   url={http://dx.doi.org/10.1038/s42254-023-00645-5},
   DOI={10.1038/s42254-023-00645-5},
   number={12},
   journal={Nature Reviews Physics},
   publisher={Springer Science and Business Media LLC},
   author={McMahon, Peter L.},
   year={2023},
   month=oct, pages={717–734} }

@article{ozkan_DNN_nature,
  title={Diffractive optical computing in free space},
  author={Hu, Jingtian and Mengu, Deniz and Tzarouchis, Dimitrios C and Edwards, Brian and Engheta, Nader and Ozcan, Aydogan},
  journal={Nature Communications},
  volume={15},
  number={1},
  pages={1525},
  year={2024},
  publisher={Nature Publishing Group UK London}
}

@article{Iran2025programmable_DNN,
  title={Programmable diffractive deep neural networks enabled by integrated rewritable metasurfaces},
  author={Zarei, Sanaz},
  journal={Scientific Reports},
  volume={15},
  number={1},
  pages={35624},
  year={2025},
  publisher={Nature Publishing Group UK London}
}

@article{optica2024holography_encrypt,
  title={Reprogrammable metasurface holographic image encryption technology based on a three-dimensional discrete hyperchaotic system},
  author={Bi, Kaiyun and Zhang, Guanmao and Zhang, Jilong and Diao, Guangchao and Xing, Bochuan and Cui, Mengjie and Ge, Zhilin and Du, Yuze},
  journal={Optics Express},
  volume={32},
  number={22},
  pages={38703--38719},
  year={2024},
  publisher={Optica Publishing Group}
}

@article{nature2025beam_steering,
  title={Metasurface based on phase change materials for electrically reconfigurable THz beam steering in copolarized transmission mode},
  author={Kumar, Krishna and Vidal, Borja and Garcia-Meca, Carlos},
  journal={Scientific Reports},
  volume={15},
  number={1},
  pages={40666},
  year={2025},
  publisher={Nature Publishing Group UK London}
}

@article{tsinghua2024onn_review,
  title={Optical neural networks: progress and challenges},
  author={Fu, Tingzhao and Zhang, Jianfa and Sun, Run and Huang, Yuyao and Xu, Wei and Yang, Sigang and Zhu, Zhihong and Chen, Hongwei},
  journal={Light: Science \& Applications},
  volume={13},
  number={1},
  pages={263},
  year={2024},
  publisher={Nature Publishing Group UK London}
}

@article{ozkan_optic_generative_nature,
  title={Optical generative models},
  author={Chen, Shiqi and Li, Yuhang and Wang, Yuntian and Chen, Hanlong and Ozcan, Aydogan},
  journal={Nature},
  volume={644},
  number={8078},
  pages={903--911},
  year={2025},
  publisher={Nature Publishing Group UK London}
}

@article{MONN_review2025programmable,
  title={Programmable metasurfaces for future photonic artificial intelligence},
  author={Abou-Hamdan, Loubnan and Marinov, Emil and Wiecha, Peter and del Hougne, Philipp and Wang, Tianyu and Genevet, Patrice},
  journal={Nature Reviews Physics},
  pages={1--17},
  year={2025},
  publisher={Nature Publishing Group UK London}
}

@article{ms_array2024singleshot_ellipsometry,
  title={Metasurface array for single-shot spectroscopic ellipsometry},
  author={Wen, Shun and Xue, Xinyuan and Wang, Shuai and Ni, Yibo and Sun, Liqun and Yang, Yuanmu},
  journal={Light: Science \& Applications},
  volume={13},
  number={1},
  pages={88},
  year={2024},
  publisher={Nature Publishing Group UK London}
}

@article{capasso2024singleshot,
  title={Metasurface-enabled single-shot and complete Mueller matrix imaging},
  author={Zaidi, Aun and Rubin, Noah A and Meretska, Maryna L and Li, Lisa W and Dorrah, Ahmed H and Park, Joon-Suh and Capasso, Federico},
  journal={Nature Photonics},
  volume={18},
  number={7},
  pages={704--712},
  year={2024},
  publisher={Nature Publishing Group UK London}
}

@article{another_segmentation_paper,
  title={Glaucoma identification using convolutional neural networks ensemble for optic disc and cup segmentation},
  author={Virbukait{\.e}, Sandra and Bernatavi{\v{c}}ien{\.e}, Jolita and Imbrasien{\.e}, Daiva},
  journal={IEEE Access},
  volume={12},
  pages={82720--82729},
  year={2024},
  publisher={IEEE}
}

@article{Zinlin2024E2Edesign,
  title={End-to-end optimization of metasurfaces for imaging with compressed sensing},
  author={Arya, Gaurav and Li, William F and Roques-Carmes, Charles and Soljacic, Marin and Johnson, Steven G and Lin, Zin},
  journal={ACS Photonics},
  volume={11},
  number={5},
  pages={2077--2087},
  year={2024},
  publisher={ACS Publications}
}

@article{e2e_framework2021design,
  title={Design framework for metasurface optics-based convolutional neural networks},
  author={Burgos, Carlos Mauricio Villegas and Yang, Tianqi and Zhu, Yuhao and Vamivakas, A Nickolas},
  journal={Applied Optics},
  volume={60},
  number={15},
  pages={4356--4365},
  year={2021},
  publisher={Optical Society of America}
}

@article{stanford2018classification,
  title={Hybrid optical-electronic convolutional neural networks with optimized diffractive optics for image classification},
  author={Chang, Julie and Sitzmann, Vincent and Dun, Xiong and Heidrich, Wolfgang and Wetzstein, Gordon},
  journal={Scientific reports},
  volume={8},
  number={1},
  pages={12324},
  year={2018},
  publisher={Nature Publishing Group UK London}
}

@article{laura_another2024information,
  title={Information-driven design of imaging systems},
  author={Pinkard, Henry and Kabuli, Leyla and Markley, Eric and Chien, Tiffany and Jiao, Jiantao and Waller, Laura},
  journal={arXiv:2405.20559},
  year={2024}
}

@article{all_optical_nonlinearity2019optica,
  title={All-optical neural network with nonlinear activation functions},
  author={Zuo, Ying and Li, Bohan and Zhao, Yujun and Jiang, Yue and Chen, You-Chiuan and Chen, Peng and Jo, Gyu-Boong and Liu, Junwei and Du, Shengwang},
  journal={Optica},
  volume={6},
  number={9},
  pages={1132--1137},
  year={2019},
  publisher={Optical Society of America}
}

@article{tsinghua2025modesign,
  title={From performance to structure: a comprehensive survey of advanced metasurface design for next-generation imaging},
  author={Zeng, Yunhui and Zhong, Haopeng and Long, Zhenwei and Cao, Hongkun and Jin, Xin},
  journal={npj Nanophotonics},
  volume={2},
  number={1},
  pages={39},
  year={2025},
  publisher={Nature Publishing Group UK London}
}

@article{goodman1978incoherent,
  title={Fully parallel, high-speed incoherent optical method for performing discrete Fourier transforms},
  author={Goodman, Joseph W and Dias, AR and Woody, LM},
  journal={Optics Letters},
  volume={2},
  number={1},
  pages={1--3},
  year={1978},
  publisher={Optical Society of America}
}

@article{capasso2014mo,
  title={Flat optics with designer metasurfaces},
  author={Yu, Nanfang and Capasso, Federico},
  journal={Nature materials},
  volume={13},
  number={2},
  pages={139--150},
  year={2014},
  publisher={Nature Publishing Group UK London}
}

@article{incoherent2023universal_linear,
  title={Universal linear intensity transformations using spatially incoherent diffractive processors},
  author={Rahman, Md Sadman Sakib and Yang, Xilin and Li, Jingxi and Bai, Bijie and Ozcan, Aydogan},
  journal={Light: Science \& Applications},
  volume={12},
  number={1},
  pages={195},
  year={2023},
  publisher={Nature Publishing Group UK London}
}

@article{zju2023knowledge_distillation,
  title={A knowledge-inherited learning for intelligent metasurface design and assembly},
  author={Jia, Yuetian and Qian, Chao and Fan, Zhixiang and Cai, Tong and Li, Er-Ping and Chen, Hongsheng},
  journal={Light: Science \& Applications},
  volume={12},
  number={1},
  pages={82},
  year={2023},
  publisher={Nature Publishing Group UK London}
}

@article{zhao2025endoscopy,
  title={Design of an infrared wide-angle metalens for medical endoscopic imaging systems},
  author={Zhao, Xinjie and Peng, Xing and Xu, Shaohui and Li, Shiqing and Cao, Hongbing and Shi, Feng},
  journal={Optics Express},
  volume={33},
  number={14},
  pages={29182--29196},
  year={2025},
  publisher={Optica Publishing Group}
}

@article{zeng2025edof,
  title={Extended depth of focus metalens toward the entire long-wave infrared spectrum through inverse design framework},
  author={Zeng, Yongcan and Ge, Xiaoling and Zhang, Yuqing and Xiao, Siyang and Zhao, Fen and Ran, Chongchong and Wu, Mingjie and Gan, Fengyuan and Wu, Jiagui and Yang, Junbo},
  journal={Optics Express},
  volume={33},
  number={18},
  pages={39081--39091},
  year={2025},
  publisher={Optica Publishing Group}
}

@article{goodam1992filtermatch,
  title={Incoherent pattern recognition with phase-only filters},
  author={van der Gracht, Joseph and Mait, Joseph N},
  journal={Optics Letters},
  volume={17},
  number={23},
  pages={1703--1705},
  year={1992},
  publisher={Optical Society of America}
}


\setcounter{figure}{0}
\setcounter{table}{0}
\setcounter{page}{0}
\setcounter{secnumdepth}{3}
\renewcommand{\thesection}{S\arabic{section}}

\renewcommand{\thefigure}{S\arabic{figure}}
\renewcommand{\thetable}{S\arabic{table}}
\renewcommand{\thepage}{S\arabic{page}}
\clearpage
\begin{center}

{\Large \textbf{Supporting Information}}

\vspace{0.8cm}

{\LARGE \textbf{Advantages of Broadband Metalenses for Generalizable Image Classification}}

\vspace{0.6cm}

{\large Yubo Zhang, Johannes Fr\"och, Jinlin Xiang, Shane Colburn, Myunghoo Lee, Zhihao Zhou, Minho Choi, Eli Shlizerman, Arka Majumdar}

\vspace{0.8cm}

{\large
\textbf{Number of pages: 13} \\
\textbf{Number of figures: 7} \\
\textbf{Number of tables: 1}
}

\end{center}
\clearpage
\section{Supplementary Material }

\subsection{Material and Fabrication of Experimental Metalens}\label{supplement: material}

\begin{figure}[H]
\centering
\includegraphics[width=0.25\linewidth]{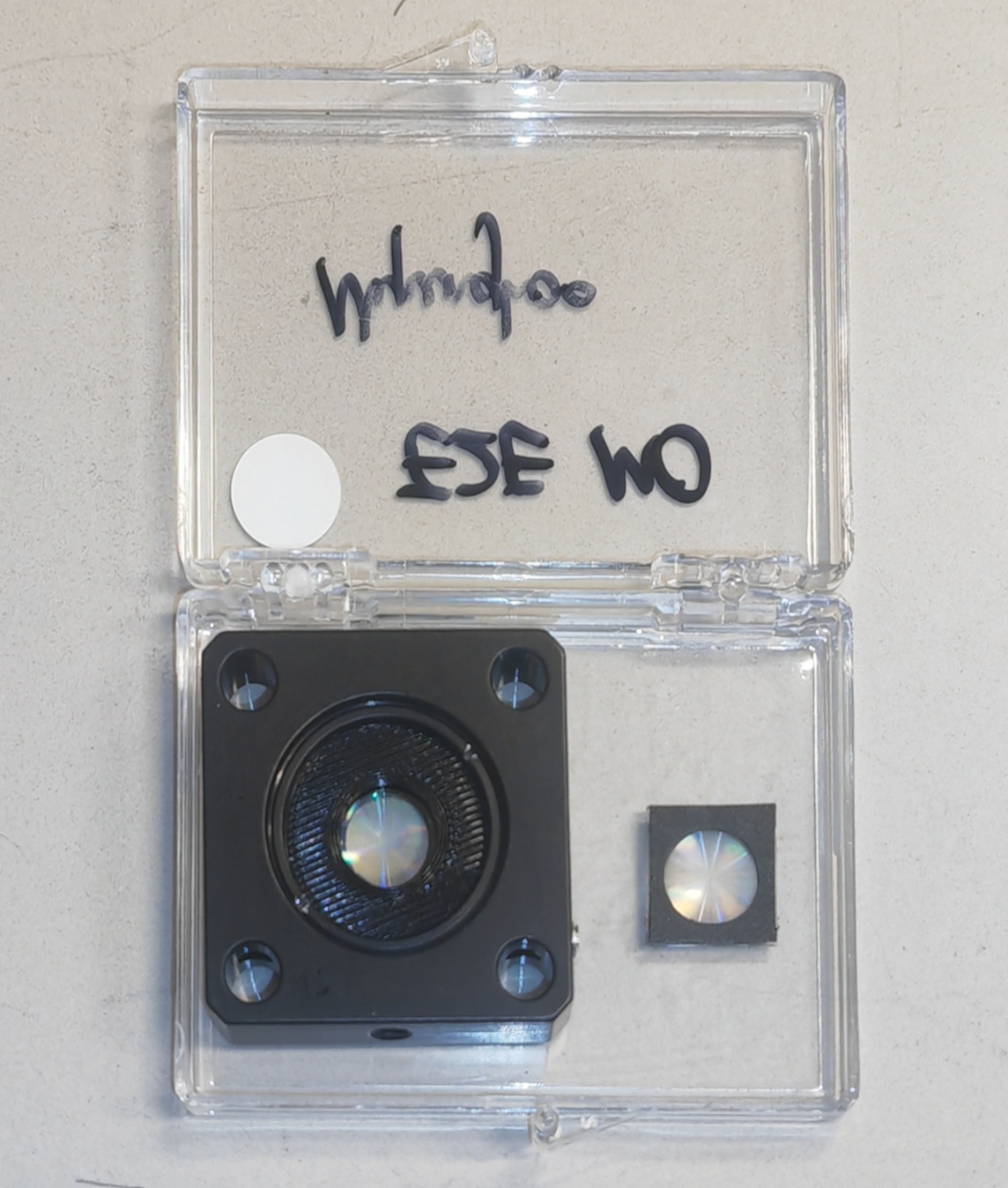}
\caption{Images of the centimeter-aperture hyperboloid (left) and broadband (right) metalenses used in the experiment. Both metalenses have identical 1 cm apertures.}
\label{fig:Mos_1cm}
\end{figure}

The metalenses (Fig.~\ref{fig:Mos_1cm}) were fabricated on a $\mathrm{Si_3N_4}$–on–quartz platform using standard nanofabrication processes. A 800nm-thick $\mathrm{Si_3N_4}$ layer was first deposited on a quartz substrate via plasma-enhanced chemical vapor deposition (PECVD), serving as the high-index metasurface layer. Following solvent cleaning and oxygen plasma treatment to improve surface adhesion, a positive electron-beam resist (ZEP 520A, 400nm thick) was spin-coated and soft-baked. A thin conductive polymer layer was subsequently applied to mitigate charging effects during exposure. The metasurface pattern, consisting of a square nanopillar array over a 1cm aperture, was defined using high-energy electron-beam lithography (100keV) with an optimized exposure dose. After development, the resist pattern was transferred into a hard mask layer, followed by inductively coupled plasma reactive ion etching using fluorine-based chemistry to etch the $\mathrm{Si_3N_4}$ layer with high anisotropy and fidelity. Residual mask and resist layers were removed through solvent cleaning and oxygen plasma ashing. The fabricated metasurfaces exhibit uniform structural quality across the centimeter-scale aperture, as verified by optical microscopy (showing radially uniform structural coloration) and scanning electron microscopy (SEM) from both top-view and tilted perspectives. The use of a single-layer $\mathrm{Si_3N_4}$ square-pillar geometry enables compatibility with scalable lithographic techniques such as nanoimprint lithography, providing a viable pathway toward large-area, high-throughput manufacturing.

\clearpage
\subsection{Definition of MTF and Its Relation to PSF and Sensor Sampling}\label{supplement: mtf_def}

The modulation transfer function (MTF) characterizes the spatial-frequency response of an imaging system and is defined as the magnitude of the optical transfer function (OTF). Here, 
$r$ denotes the spatial coordinate in the image (or sensor) plane, and 
$f$ denotes the corresponding spatial-frequency coordinate. The OTF is given by the Fourier transform of the point spread function (PSF),
\[
\mathrm{OTF}(\mathbf{f}) = \mathcal{F}\{\mathrm{PSF}(\mathbf{r})\},
\]
and the MTF is defined as
\[
\mathrm{MTF}(\mathbf{f}) = \frac{|\mathrm{OTF}(\mathbf{f})|}{|\mathrm{OTF}(\mathbf{0})|},
\]
such that the zero-frequency value is normalized to unity.

This formulation establishes a direct correspondence between the spatial-domain response and its frequency-domain representation: fine spatial features in the PSF correspond to higher spatial-frequency components in the MTF, while broader PSFs lead to attenuation of high-frequency content.

In practical imaging systems, the sensor introduces additional filtering due to finite pixel size. Each pixel integrates the incident intensity over its area, which can be modeled as a convolution with a rectangular kernel in the spatial domain,
\[
I_{\mathrm{meas}}(\mathbf{r}) = I(\mathbf{r}) * \mathrm{rect}\left(\frac{\mathbf{r}}{p}\right),
\]
where \( p \) denotes the pixel pitch. In the frequency domain, this corresponds to multiplication by a sinc function,
\[
H_{\mathrm{sensor}}(\mathbf{f}) = \mathrm{sinc}(p f_x)\,\mathrm{sinc}(p f_y),
\]
resulting in attenuation of high spatial frequencies.

In addition to this averaging effect, discrete sampling imposes a Nyquist limit on the order of \( 1/(2p) \), beyond which spatial-frequency components cannot be reliably captured. The effective system response is therefore jointly determined by the optical MTF and the sensor transfer function, and only spatial-frequency components within this passband are transmitted to the digital backend.

Pixel binning further increases the effective pixel size, leading to stronger spatial averaging and a narrower detectable frequency band. As a result, sensor sampling defines the spatial-frequency passband through which information from the optical front end is conveyed to subsequent processing stages.

For multi-wavelength analysis, we note that the MTF of each color channel is normalized to its zero-frequency value and is therefore invariant to overall scaling of the corresponding PSF. Consequently, differences in total throughput across wavelengths do not affect the MTF. Cross-channel comparisons thus reflect differences in the distribution of spatial-frequency transmission rather than differences in absolute intensity.

\clearpage
\subsection{Experimental Setup and Measurement}\label{supplement: exp_setup}

\begin{figure}[H]
\centering
\includegraphics[width=\linewidth]{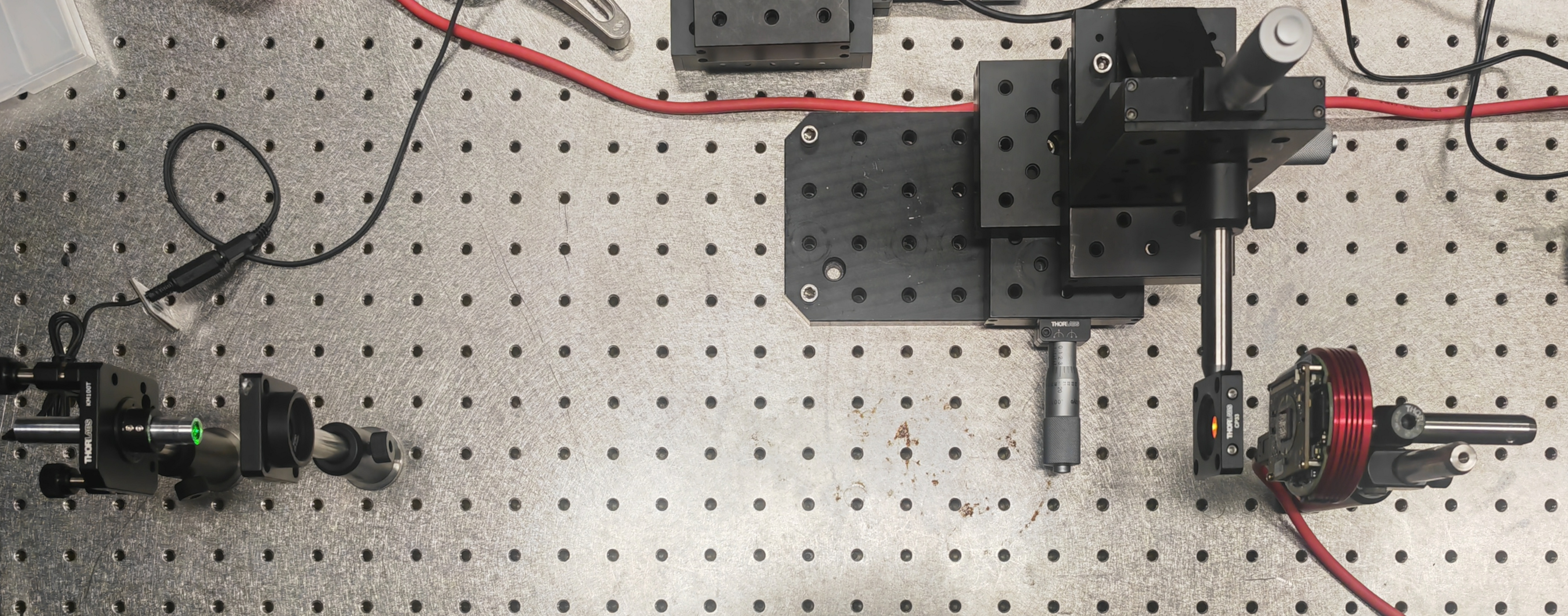}
\caption{Experimental setup for PSF measurement (schematic, not to scale).}
\label{fig:Exp_setup}
\end{figure}

Figure~\ref{fig:Exp_setup} illustrates the experimental setup used to measure the point spread functions (PSFs) of the two metalenses. Single-wavelength laser sources (red, green, and blue; Thorlabs) were used sequentially for illumination. The beams were spatially filtered using a 25$\mu$m pinhole (Thorlabs) and propagated over a distance of approximately 1m to approximate collimated illumination at the meta-optics. The resulting PSFs were recorded at the focal plane using a ZWO ASI678MC camera with a pixel size of 2$\mu$m.

For ImageNet30 data acquisition, a Dell monitor (1920$\times$1080 resolution) was positioned at the object plane to display input images. Each image occupied a $900\times900$ pixel region on the monitor. The sensor was placed at the focal plane of the metalens (focal length 2cm). The object distance was adjusted such that each image formed a $4.000\text{mm} \times 4.000\text{mm}$ region on the sensor, corresponding to an effective field of view of approximately $11.4^\circ$.

An automated acquisition pipeline was used to capture approximately 36,000 encoded images from ImageNet30 for each metalens. The camera exposure time was dynamically adjusted to avoid saturation and maintain consistent signal levels. All captured images were subsequently processed with spatial binning to simulate different effective sensor pixel sizes.

\clearpage

\subsection{More Imaging Results with the Centimeter-Aperture Metalenses on Imagenet30 Samples}\label{supplement: more imaging results}

\begin{figure}[H]
\centering
\includegraphics[width=\linewidth]{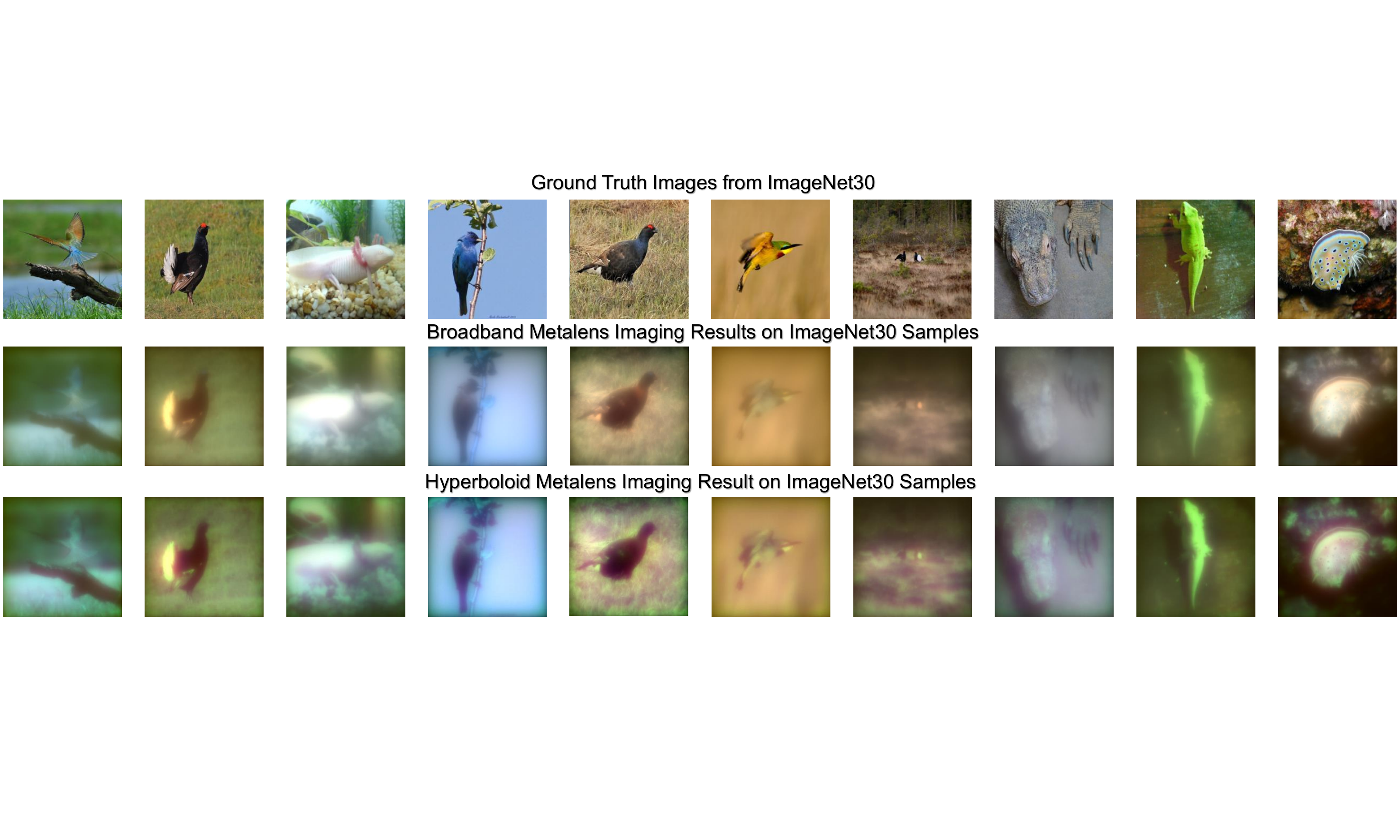}
\caption{Comparison between the broadband metalens and the hyperboloid metalens with identical imaging condition. 
Top: Groud truth samples from Imagenet30. 
Middle: Broadband metalens imaging results, exhibiting improved color fidelity and smoother tonal transitions. 
Bottom: Hyperboloid metalens imaging results, showing sharper contours but reduced color fidelity due to chromatic aberrations.
}
\label{fig:imgs_e2e_vs_hyper}
\end{figure}

We compare the broadband metalens with a baseline hyperboloid metalens of the same aperture and $f$-number. Supplement Fig.~\ref{fig:imgs_e2e_vs_hyper} shows (top) the digital ground-truth reference, (middle) the broadband metalens imaging outputs, and (bottom) the hyperboloid lens imaging results. Qualitatively, the broadband metalens preserves higher color fidelity and smoother global intensity gradients across the field, with reduced inter-channel crosstalk and more accurate saturation in chromatic regions. In contrast, the hyperboloid lens yields crisper contours and stronger high-frequency edge contrast, but at the expense of color accuracy—exhibiting desaturation and mild hue shifts consistent with residual broadband chromatic aberrations and channel-dependent MTF. Overall, the broadband metalens provides more faithful color reproduction, whereas the hyperboloid lens emphasizes geometric detail (edges/contours). These qualitative trends are consistent across the dataset and align with our system characterization.

\clearpage

\subsection{Classification Performance of Fabricated Metalenses under Large-Pixel Sensor Regimes}

To further investigate the sensor-limited regime, we evaluate classification performance at large effective sensor pixel sizes using experimentally measured images. Different pixel configurations are simulated by spatially binning the high-resolution measurements, covering pixel sizes from approximately 60~$\mu$m to 2000~$\mu$m (corresponding to effective resolutions from 67$\times$67 to 2$\times$2).

As the pixel size increases, the sensor averages the incident optical intensity over a larger area, suppressing high-spatial-frequency content and imposing a stronger low-pass constraint on the captured signal.

\begin{figure}[H]
\centering
\includegraphics[width=\linewidth]{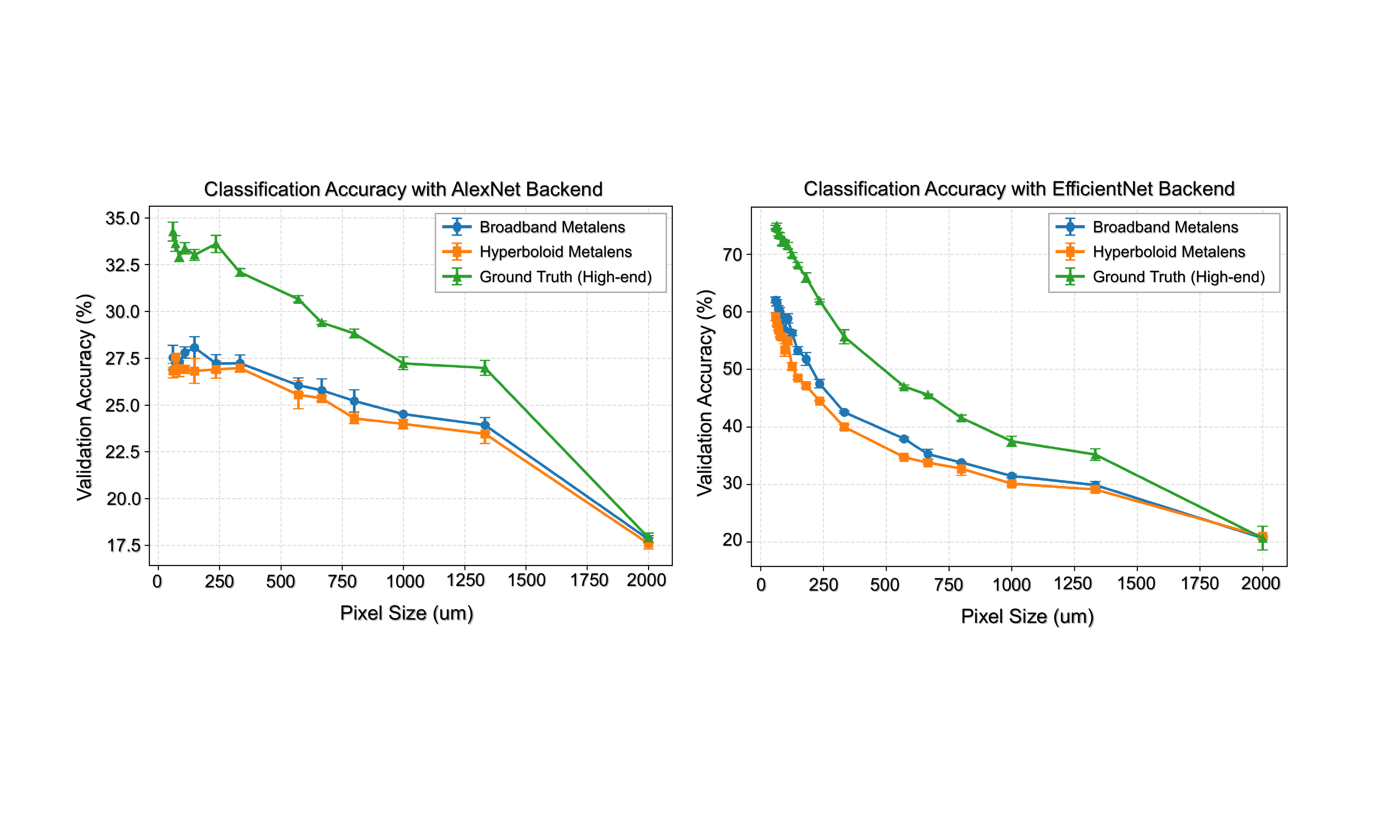}

\caption{
Classification accuracy as a function of effective sensor pixel size for AlexNet and EfficientNet backends using experimentally measured images. Increasing pixel size reduces classification accuracy and progressively diminishes the performance differences between optical systems, with convergence observed in the large-pixel, strongly sensor-limited regime. Error bars denote one standard deviation across three independent runs with different random seeds, and the reported accuracy for each run corresponds to the best validation accuracy achieved during training.
}
\label{fig: large_pixel_regime}
\end{figure}

As shown in Fig.~\ref{fig: large_pixel_regime}, the classification accuracy of all optical encoders decreases as the pixel size increases. Notably, the performance gap between different optical systems also diminishes at large pixel sizes. At 2000~$\mu$m, the metalens achieves classification accuracy comparable to that of the high-end reference optics.

Although such extremely low-resolution settings are not practically useful for classification by themselves, they serve to isolate the sensor-limited regime and help illustrate the system behavior in this limit. In particular, they support our conclusion that, once the sensor-imposed MTF cutoff becomes sufficiently restrictive, the advantage of high-end refractive optics over metalenses is substantially reduced.

This behavior is consistent with the frequency-domain interpretation in the main text: when sensor resolution is low, only spatial-frequency components within the sensor-detectable band contribute meaningfully to the task, making differences in higher-frequency optical performance less important. This further highlights the importance of in-band spatial-frequency utilization in sensor-limited computer vision systems.

\clearpage

\subsection{Limited Discriminability of Metalens Encodings in Standard Feature Spaces}\label{supplement:clustering}

To assess how different meta-optical encoders preserve dataset information, we analyze feature distributions at the sensor plane using a frozen ImageNet-pretrained ResNet-50 encoder. Features extracted from experimentally captured images (broadband metalens, hyperboloid metalens, and digital ground truth) are clustered using $k$-means, and performance is evaluated using Normalized Mutual Information (NMI) as a function of the number of clusters $K$.

\begin{figure}[H]
\centering
\includegraphics[width=\linewidth]{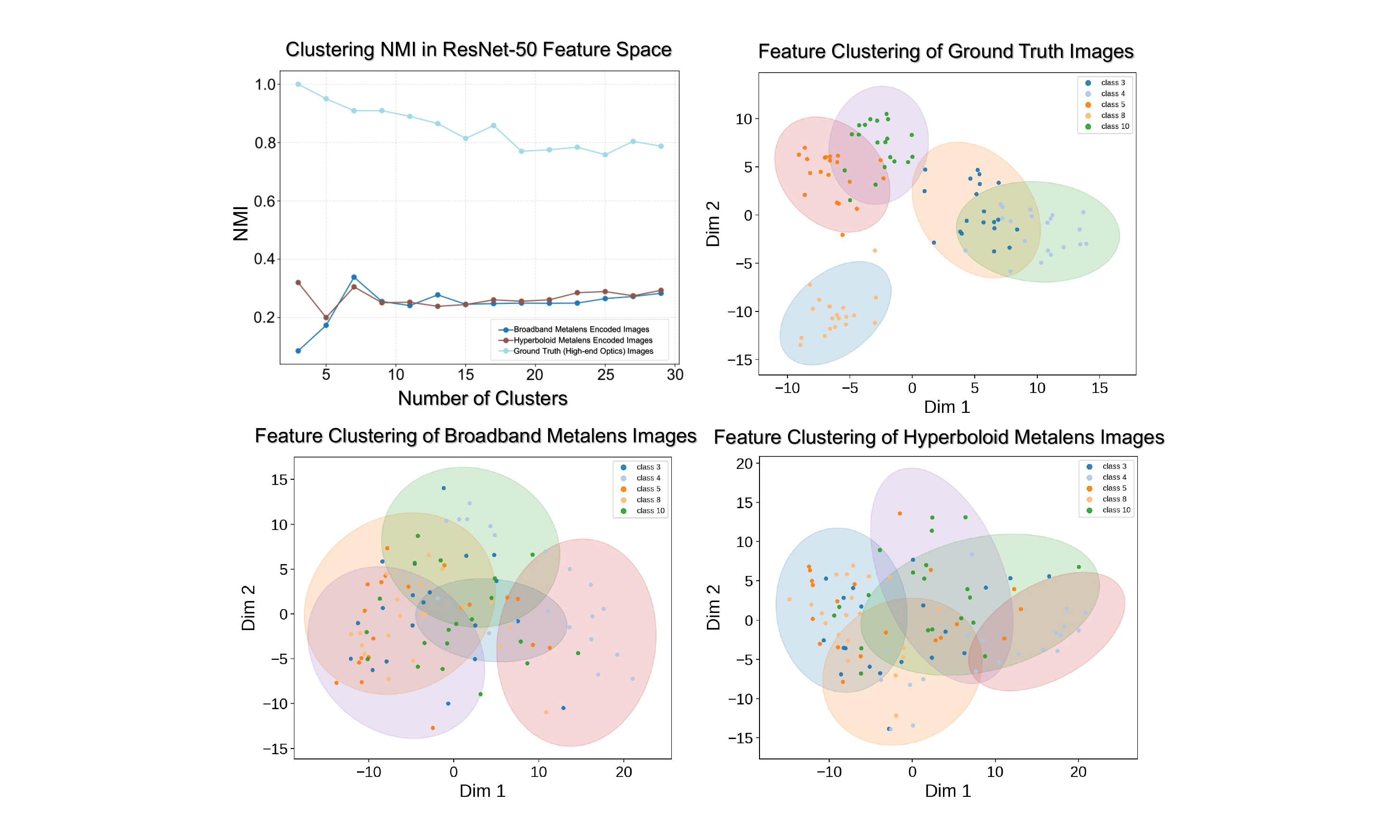}
\caption{(Upper left) Normalized Mutual Information (NMI) versus number of clusters for the broadband metalens and the hyperboloid metalens, showing similar trends. 
(Other panels) 2D visualization of clustered features for the first five classes under three optics: (Upper right) digital ground-truth(high-end optics) images (NMI = 1.000); (Lower left) broadband metalens encoded images (NMI = 0.214); (Lower right) hyperboloid metalens encoded images (NMI = 0.282). 
All results are obtained at the sensor plane, reflecting the information preserved by different optical encoders prior to the digital backend.
}

\label{fig:supplement_clustering}
\end{figure}

As shown in Fig.~\ref{fig:supplement_clustering}, the two optical encoders yield broadly similar NMI curves, with the hyperboloid lens slightly higher at larger $K$. This suggests that, in a standard deep feature space, the class-separable structure of the sensor-plane measurements is comparable across the two front ends.

These results indicate that differences between the two encoding strategies are not clearly distinguishable in either feature space or visual inspection, as sharper edges or higher contrast do not necessarily translate to improved task performance. This motivates the use of physically grounded, frequency-domain metrics—such as the MTF to more directly characterize the information transmitted by the optical front end.

\clearpage

\subsection{Methods of End-to-end Optimization}\label{supplement:e2e}

We consider three wavelengths (red, green, and blue). The trainable optical variables are the rotationally symmetric phase coefficients at the green wavelength, parameterized as
\[
\phi_G(r) \;=\; \sum_{k={0,2,4,\dots}}^{K} c_k\, r^k,\quad r\in[0,1],
\]
where $r$ is the normalized pupil radius and $\mathbf{c}=\{c_k\}$ are optimized during training. In practice, we use $K=16$ (8 coefficients with even orders) and initialize the phase profile as a hyperboloid.

To enforce physically consistent wavelength-dependent behavior, we use polynomial surrogate models fitted to RCWA simulations to relate phase and scatterer geometry. Specifically, an approximate inverse mapping $s(r)\approx \mathcal{F}_G^{-1}(\phi_G(r))$ is used to infer the meta-atom geometry from the green phase, and the corresponding red and blue phases are obtained via forward mappings $\phi_\lambda(r)\approx \mathcal{F}_\lambda(s(r))$ for $\lambda\in\{R,B\}$. These mappings are implemented as differentiable polynomial functions to enable gradient-based optimization.

Given $\{\phi_\lambda\}$, wavelength-dependent PSFs $h_\lambda(d;\mathbf{c})$ are computed using a scalar diffraction model based on the angular spectrum method. The optical system has a focal length of 1mm and a circular aperture of 0.5mm diameter. 

The sensor-plane image is modeled as a shift-invariant convolution within the isoplanatic regime:
\[
y_\lambda \;=\; x_\lambda \;\ast\; \mathcal{S}_{d}\!\left(h_\lambda(d;\mathbf{c})\right),\qquad
y \;=\; \sum_{\lambda\in\{R,G,B\}} \gamma_\lambda\, y_\lambda \;+\; \eta,
\]
where $\mathcal{S}_{d}$ accounts for geometric scaling with object distance, $\gamma_\lambda$ denotes spectral weighting (set to unity), and $\eta$ models sensor noise (Poisson and Gaussian).

The rendered image $y$ is fed into a digital backbone $f_\theta(\cdot)$ (e.g., EfficientNet-B0, ResNet18, or AlexNet) to produce classification logits. We jointly optimize the optical coefficients and network parameters using a cross-entropy loss:
\[
\min_{\mathbf{c},\,\theta}\; \frac{1}{N}\sum_{i=1}^{N} \mathcal{L}\!\big(f_\theta(y_i(\mathbf{c})),\, t_i\big)\;+\;\lambda_{\text{reg}}\,\mathcal{R}(\mathbf{c}),
\]
where $\mathcal{R}$ is an optional regularization term (e.g., coefficient magnitude constraints).

During training, we use the Adam optimizer with a learning rate of $10^{-4}$ and jointly update optical and digital parameters via backpropagation. Different effective learning rates are applied by scaling the phase coefficients. The input images are spatially downsampled using adaptive average pooling (e.g., $70\times70$) to simulate different sensor resolutions.

Training is performed for up to 100 epochs with early stopping (patience of 20 epochs). The reported performance corresponds to the best validation accuracy achieved during training.

\clearpage
\subsection{Transferability of End-to-End Optimized Optical Encoders Across Sensors and Backends}\label{supplement: e2e_transfer}

To further examine the transferability of end-to-end optimized optical encoders, we conduct additional experiments across varying sensor resolutions and digital backends. In these experiments, the optical encoder is first optimized under a specific training configuration (i.e., a fixed sensor resolution and digital backend) and then frozen (using a 24-parameter radial phase polynomial representation). The frozen encoder is subsequently evaluated under mismatched target settings, where only the digital backend is retrained.

\begin{figure}[H]
\centering
\includegraphics[width=\linewidth]{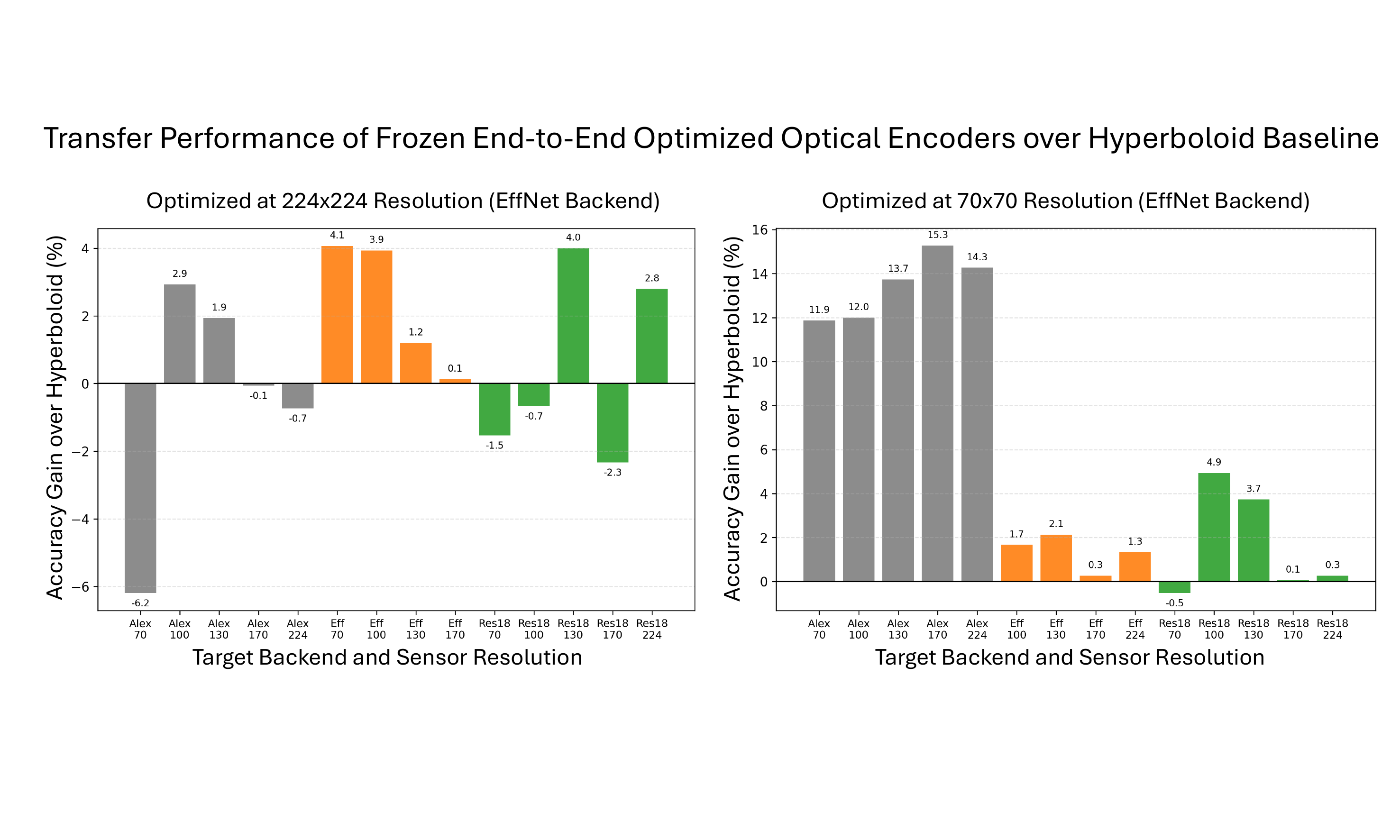}

\caption{Transfer performance of frozen end-to-end optimized optical encoders across different sensor resolutions and digital backends. In each panel, the optical encoder is first optimized end-to-end together with an EfficientNet digital backend under the indicated sensor resolution, and then frozen, while only the target digital backend is retrained for transfer evaluation. The left and right panels correspond to source sensor resolutions of \(224\times224\) and \(70\times70\), respectively. Within each panel, gray, orange, and green bars denote transfer results with AlexNet, EfficientNet, and ResNet backends with different sensor resolutions, respectively. Bars show the accuracy gain over the hyperboloid baseline.
Encoders optimized at lower resolution (right) generally retain positive gains when transferred to higher-resolution settings, whereas high-resolution-optimized encoders (left) exhibit reduced or inconsistent gains when applied to lower-resolution configurations.
}

\label{fig:supplement_e2e_general}
\end{figure}

Figure~\ref{fig:supplement_e2e_general} summarizes the performance difference between the frozen end-to-end optimized encoder and the hyperboloid baseline across different target configurations. While the optimized encoder outperforms the baseline under the original training setting, its relative advantage varies across sensor resolutions and backend architectures, and can in some cases be reduced or reversed.

Interestingly, we observe an asymmetric transfer behavior across sensor resolutions. Encoders optimized under lower-resolution settings can often transfer to higher-resolution configurations while retaining a performance advantage over the baseline. In contrast, encoders optimized at higher resolutions generally do not exhibit the same level of transferability when applied to lower-resolution settings.

This behavior is consistent with the frequency-domain interpretation discussed in the main text. At lower resolutions, end-to-end optimization tends to prioritize spatial-frequency transmission within a more restricted sensor-detectable band. The resulting MTF redistribution remains largely within the passband of higher-resolution sensors and can therefore still be utilized. Conversely, high-resolution-optimized encoders allocate more capacity to higher spatial frequencies that are not accessible under lower-resolution sampling, leading to reduced effectiveness when transferred.

These results suggest that end-to-end optimized optical encoders are sensitive to the training configuration and may not transfer uniformly across different imaging conditions. This observation is consistent with the discussion in the main text and highlights a potential limitation of fully task-specific optical optimization in hybrid ONNs.

\clearpage

\subsection{MTF Evolution of End-to-End Optimized Metalenses with Different Digital Backends}\label{supplement: acc_table}

We perform end-to-end simulations on the ImageNet30 classification task using a hyperboloid phase profile as both the initialization and baseline, and evaluate the corresponding changes in MTF integrals and classification accuracy across different digital backends.

All optimized designs consistently achieve higher classification accuracy when the MTF distribution becomes more balanced across the RGB channels, particularly when the green-channel contribution is redistributed relative to the red and blue channels.

We further observe that higher-capacity backend models (e.g., ResNet) more effectively utilize the transmitted spatial-frequency information, resulting in larger end-to-end validation accuracy gains compared with lower-capacity models (e.g., AlexNet).

All these observations are consistent with the trends reported in the main text and further support the interpretation that frequency-domain information preservation contributes to ONN performance.

\begin{table}[htbp]
\centering
\renewcommand{\arraystretch}{1.3} 
\caption{Comparison of accuracy and MTF metrics across different Digital backends.}
\begin{tabular}{lcccccccc}
\toprule
                                   & Alexnet & Effnetb0 & Effnetb7 &  Resnet18 & Resnet50 & vgg11 & vgg19\\
\midrule
Digital accuracy(\%)               & 37.80   & 77.60 & 77.60 & 67.53 & 70.47 & 69.40 & 69.33\\
Hyperboloid simulated accuracy(\%) & 28.27   & 64.60 & 65.93 & 52.93 & 56.87 & 54.33 & 52.73\\
End-to-end simulated accuracy(\%)  & 29.87   & 66.73 & 67.13 & 56.40 & 60.53 & 55.27 & 57.00\\
Red MTF integral improvement(\%)   & 618.89  & 56.61 & 75.31 & 305.02 & 34.03 & 1208.26 & 184.97\\
Green MTF integral improvement(\%) & -65.41  & -59.09 & -84.09 & -43.53 & -46.60 & -85.46 & -89.61\\
Blue MTF integral improvement(\%)  & 105.75  & 136.49 & 292.00 & 46.76 & 120.22 & 260.45 & 181.18\\
\bottomrule
\end{tabular}

\label{tab:effnet_comparison}
\end{table}

\clearpage

\subsection{Details of the Synthesized MTF Experiment}\label{supplement: synmtf}

To systematically investigate the role of frequency-domain transmission, we construct a set of synthetic OTFs that emulate meta-optical encoders with controlled spatial-frequency responses.

Given an RGB input image $x \in \mathbb{R}^{H \times W \times 3}$ (with $H=W=224$), we transform each channel into the Fourier domain:
\[
X_\lambda(f_x,f_y) = \mathcal{F}\{x_\lambda\}, \quad \lambda \in \{R,G,B\}.
\]

We define a normalized radial spatial-frequency coordinate:
\[
f_r = \sqrt{f_x^2 + f_y^2},
\]
where $(f_x,f_y)$ are expressed in units of cycles/pixel and normalized such that the Nyquist frequency corresponds to $f_r = 0.5 cycles/pixel$.

For each channel, we construct a radially symmetric OTF of Gaussian form:
\[
\mathrm{OTF}_\lambda(f_r) = \exp\!\left[-\left(\frac{f_r}{f_{\mathrm{spread},\lambda}}\right)^2\right],
\]
where $f_{\mathrm{spread},\lambda}$ controls the effective bandwidth of the channel.

The filtered image is then obtained via:
\[
y_\lambda = \mathcal{F}^{-1}\left( X_\lambda \cdot \mathrm{OTF}_\lambda \right),
\quad
y = \mathrm{concat}(y_R, y_G, y_B).
\]

This procedure enables independent control over spatial-frequency transmission in each color channel while preserving the overall image structure.

\textbf{Parameter sweep.}
We sample the bandwidth parameters as:
\[
f_{\mathrm{spread},\lambda} \in \{0.10,\,0.15,\,0.25,\,0.35,\,0.45\},
\]
resulting in $5^3 = 125$ distinct RGB frequency-response configurations.

\subsubsection{In-band MTF Metrics}

To emulate sensor-limited imaging, we define an in-band cutoff frequency $f_0$ (e.g. 0.2 in cycles/pixel). All MTF-based quantities are evaluated within the region:
\[
f_r \leq f_0.
\]

The in-band MTF for each channel is defined as:
\[
\mathrm{MTF}_\lambda^{\mathrm{inband}} = 
\frac{1}{|\Omega|}
\int_{f_r \leq f_0} \left| \mathrm{MTF}_\lambda(f_r) \right| \, df,
\]
where $\Omega$ denotes the in-band region and the normalization ensures comparability across configurations.

We further define:

\textbf{(1) Wavelength-averaged in-band MTF:}
\[
\overline{\mathrm{MTF}}^{\mathrm{inband}} = 
\frac{1}{3} \sum_{\lambda \in \{R,G,B\}} 
\mathrm{MTF}_\lambda^{\mathrm{inband}}.
\]

\textbf{(2) RGB imbalance:}
\[
\mathrm{Imbalance} = 
\sqrt{\frac{1}{3} \sum_{\lambda}
\left( \mathrm{MTF}_\lambda^{\mathrm{inband}} - 
\overline{\mathrm{MTF}}^{\mathrm{inband}} \right)^2 }.
\]

These two quantities respectively characterize the overall amount of preserved spatial-frequency information and its distribution across color channels.

\subsubsection{Digital Backend and Training Protocol}

The filtered images are used as inputs to a fixed digital backbone (EfficientNet-B0). To isolate the effect of optical filtering, only the final classification layer is trained on ImageNet30, while the backbone weights remain frozen.

For each synthetic configuration, the model is trained using cross-entropy loss. Early stopping is applied if the validation accuracy does not improve for 20 consecutive epochs. The reported accuracy corresponds to the best validation performance during training.

\subsubsection{Additional Metrics}

In addition to the in-band MTF and RGB imbalance, we evaluate several alternative frequency-domain descriptors, including bandwidth measures (e.g., half-maximum cutoff), spectral centroid, and channel-wise energy concentration. These metrics exhibit qualitatively consistent trends with respect to classification accuracy, further supporting the robustness of the observed correlations.

\begin{figure}[H]
\centering
\includegraphics[width=\linewidth]{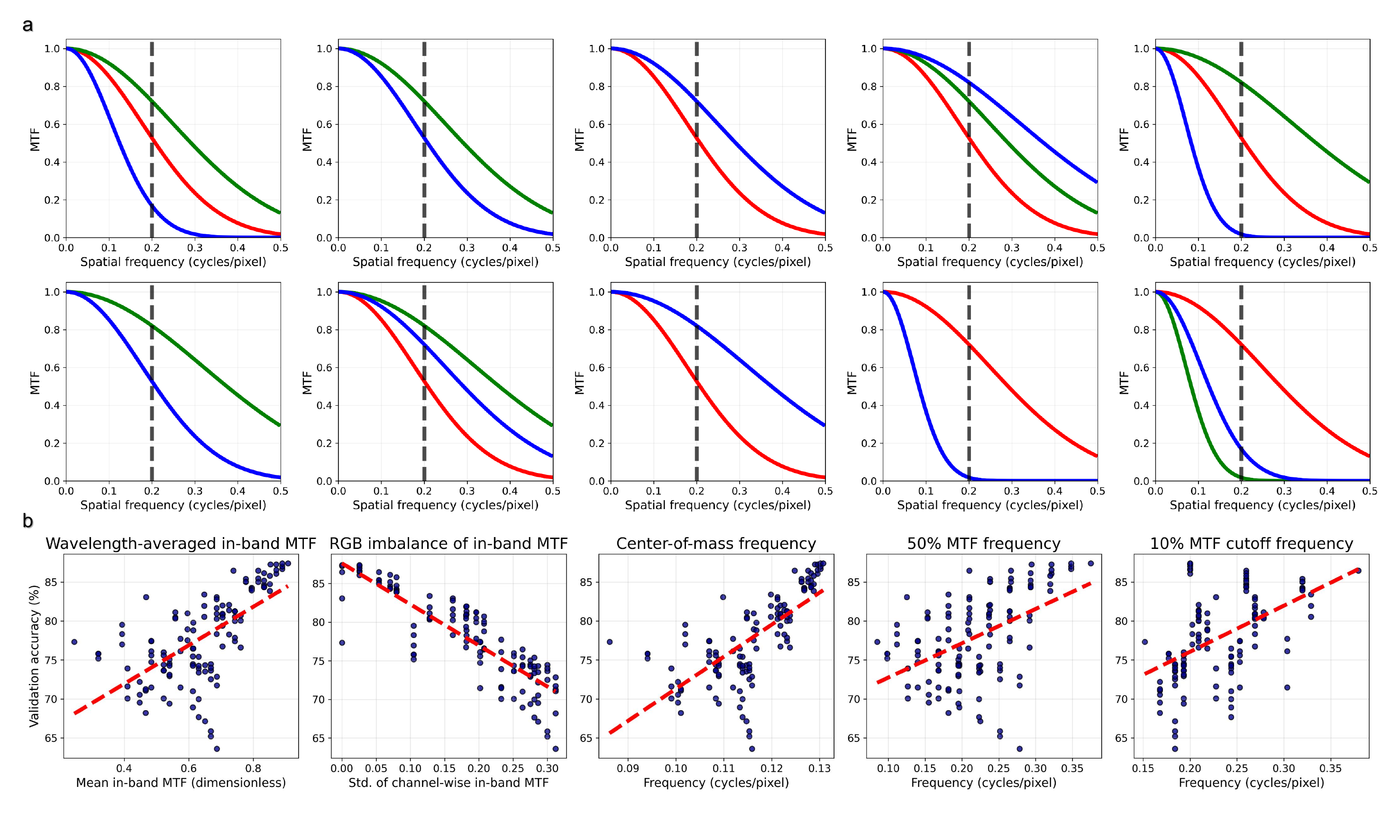}
\caption{a. Representative examples of radially symmetric RGB MTFs used to emulate meta-optical encoders with varying bandwidths and cross-channel balance. The vertical dashed line indicates the sensor cutoff frequency $f_0$, defining the in-band region.
b. Relationship between validation accuracy and different frequency-domain metrics across all synthesized configurations. 
Accuracy versus wavelength-averaged in-band MTF shows a positive correlation (Pearson correlation coefficient $r=0.632$), indicating that greater preservation of spatial-frequency content within the detectable band is associated with improved performance. 
Accuracy versus RGB imbalance (standard deviation of in-band MTF across channels) shows a strong negative correlation ($r=-0.867$), suggesting that cross-channel imbalance is detrimental. 
Accuracy versus center-of-mass frequency exhibits a moderate positive correlation ($r=0.703$). 
Alternative bandwidth-based metrics, including the 50\% MTF frequency ($r=0.481$) and 10\% cutoff frequency ($r=0.471$), show weaker but consistent trends. 
Overall, we find that classification performance is most consistently associated with the amount and spectral balance of in-band spatial-frequency transmission.
}
\label{fig:supplement_synmtf}
\end{figure}


\bibliographystyle{achemso}

\end{document}